\documentclass[draftclsnofoot, 12pt, onecolumn, journal]{IEEEtran}

\usepackage{graphicx}
\usepackage{color}
\usepackage{amsmath}
\usepackage{amsfonts}
\usepackage{epsfig}
\usepackage{epstopdf}
\usepackage{amssymb}
\graphicspath{{./Figures/}}

\usepackage{paralist}

\usepackage[tight,footnotesize]{subfigure}
\usepackage[noadjust]{cite}

\newfont{\bbb}{msbm10 scaled 500}

\newfont{\bb}{msbm10 scaled 1100}

\newcommand{\gv}{{\bf g}}
\newcommand{\hv}{{\bf h}}

\newcommand{\pv}{{\bf p}}

\newcommand{\wv}{{\bf w}}
\newcommand{\vv}{{\bf v}}
\newcommand{\xv}{{\bf x}}

\newcommand{\zv}{{\bf z}}

\newcommand{\Am}{{\bf A}}

\newcommand{\Id}{{\bf I}}

\newcommand{\Pm}{{\bf P}}

\newcommand{\Wm}{{\bf W}}

\newcommand{\Ym}{{\bf Y}}
\newcommand{\Zm}{{\bf Z}}

\newcommand{\Kc}{{\cal K}}

\newcommand{\alphav}{\hbox{\boldmath$\alpha$}}

\newtheorem{theorem}{Theorem}

\definecolor{OXO-emph}{RGB}{153,0,0}

\newcommand{\plusminus}{%
  \mathrel{%
    \vcenter{\offinterlineskip
      \ialign{##\cr$-$\cr\noalign{\kern-1.5pt}$+$\cr}%
    }%
  }%
}

\allowdisplaybreaks

\begin{document}
%
\title{Vulnerabilities of Massive MIMO Systems Against Pilot Contamination Attacks}

\author{\IEEEauthorblockN{Berk~Akgun, Marwan~Krunz, and O.~Ozan~Koyluoglu}
\thanks{An abridged version of this paper will appear in the IEEE CNS 2017 Conference, Las Vegas, October 9-11, 2017.}
\thanks{This research was supported in part by the National Science Foundation (grants $\#$ CNS-1409172, CNS-1513649, IIP-1265960, and CNS-1617335) and by Qatar Foundation (grant $\#$ NPRP 8-052-2-029). Any opinions, findings, conclusions, or recommendations expressed in this paper are those of the author(s) and do not necessarily reflect the views of the NSF and QF.}
\thanks{The authors are with the Department
of Electrical and Computer Engineering, University of Arizona, Tucson,
AZ, 85721 USA
(email: berkakgun@email.arizona.edu, ozan@email.arizona.edu, krunz@email.arizona.edu).}
}

\maketitle
\thispagestyle{empty}
\pagestyle{empty}

\begin{abstract}
We consider a single-cell massive MIMO system in which a base station (BS) with a large number of antennas transmits simultaneously to several single-antenna users in the presence of an attacker. The BS acquires the channel state information (CSI) based on uplink pilot transmissions. In this work, we demonstrate the vulnerability of CSI estimation phase to malicious attacks. For that purpose, we study two attack models. In the first model, the attacker aims at minimizing the sum-rate of downlink transmissions by contaminating the uplink pilots. In the second model, the attacker exploits its in-band full-duplex capabilities to generate jamming signals in both the CSI estimation and data transmission phases. We study these attacks under two downlink power allocation strategies when the attacker knows and does not know the locations of the BS and users. The formulated problems are solved using stochastic optimization, Lagrangian minimization, and game-theoretic methods. A closed-form solution for a special case of the problem is obtained. Furthermore, we analyze the achievable individual secrecy rates under a pilot contamination attack, and provide an upper bound on these rates. Our results indicate that the proposed attacks degrade the throughput of a massive MIMO system by more than $50\%$.
\end{abstract}

\begin{IEEEkeywords}
Massive MIMO, pilot contamination, physical layer security, active attack, stochastic optimization.
\end{IEEEkeywords}

\maketitle

\section{Introduction}

Massive multiple-input multiple-output (MIMO) is one of the key technologies in the upcoming 5G systems.
It is envisioned that a cellular base station (BS) in 5G systems will be equipped with a very large antenna array, e.g., hundreds of antennas or more, boosting
the transmission rate by orders of magnitude compared to conventional MIMO systems.
Even though MIMO is a well-studied concept in wireless communications, massive MIMO requires novel techniques to overcome new design challenges, and as such it has received significant attention from researchers over the last few years (see, for example, \cite{Bj2016}, \cite{Larsson2014}, \cite{Lu2014}, and the references therein).

One of the important issues in massive MIMO systems is pilot contamination (PC) \cite{Marzetta2010}.
Because of the large number of antennas at the BS and the relatively short channel coherence time, the channel state information (CSI) between the BS and various users must be estimated frequently using uplink pilot transmissions.
Assuming channel reciprocity, the BS utilizes these CSI estimates for downlink data transmissions.
However, due to the limited number of orthogonal pilot sequences (e.g., in the order of tens \cite{Marzetta2010}), users in neighboring cells may share the same pilots.
Interference among these pilots causes erroneous CSI estimates at the BS, leading to poor system performance. 

In \cite{Zhou2012} the authors studied an attack that exploits vulnerabilities of the channel training phase  in time division duplexing (TDD) systems.
The key idea behind this attack is to contaminate uplink pilot transmissions and cause an erroneous uplink channel estimation.
Typically, if the CSI is available, the BS would use MIMO beamforming techniques such as maximum-ratio transmission (MRT) to maximize the signal-to-noise-ratio (SNR) at users.
However, the benefits of these techniques vanish rapidly if the CSI estimates are erroneous.
A self-contamination technique was proposed in \cite{Tugnait2015} to detect this type of attack.
The authors in \cite{Kapetanovic2015} proposed another approach in which the legitimate user transmits four random phase-shift keying symbols, and the BS checks the correlation matrix of the received signals.
Based on the ratio of two largest eigenvalues of this matrix, the BS detects the attack. 
Secure transmissions for TDD-based massive MIMO systems was studied in \cite{Wu2015} in the presence of an active eavesdropper.
The authors derived the optimal power allocation for the information and artificial noise (AN) signals at the BS such that secrecy is asymptotically guaranteed, i.e., as the number of BS antennas ($M$) tends to infinity.
In \cite{Basciftci}, the authors proposed providing secrecy against PC attacks by keeping pilot assignments hidden and using a pilot set that scales with $M$.
However, there are two main problems with this scheme.
First, it requires a longer pilot transmission phase, which increases the overhead and decreases the throughput.
Second, computationally intensive cryptographic methods are required to keep pilot assignments hidden.
All of the papers discussed above consider an attacker that targets one user at a time.
Even when a multiuser system is in place, the attacker randomly selects a given user and contaminates its pilot sequence.
Given that one of the key aspects of massive MIMO systems is to serve tens of users simultaneously, the vulnerabilities of these systems to a multiuser pilot contamination attack should be investigated.

In this paper, we consider a single-cell multiuser massive MIMO network in the presence of an attacker.
We study two attack models.
In the first model, the attacker aims at minimizing the sum-rate of downlink transmissions by contaminating uplink pilot transmissions.
We derive the downlink transmission rates with and without the PC attack by exploiting the \textit{channel hardening effect} (effect of small-scale fading on channel gains vanishes as $M$ tends to infinity) in massive MIMO.
Optimal attack strategies are then investigated for two different cases: when the attacker knows the locations of the BS and users and when she does not have this information. 
Considering a fixed power allocation strategy for downlink data transmissions, convex problems are formulated for the optimal PC attack.
These problems are solved via the interior-point and Lagrangian minimization methods.
We obtain a closed-form solution for the case of perfect information, i.e., known topology at the attacker.
This solution represents a lower bound on the downlink sum-rate of massive MIMO systems under an optimal PC attack and a fixed BS transmission power.
Then, we study the scenario where the BS optimizes its own power allocation scheme in the presence of PC attacks.
For this case, a game-theoretic problem formulation is considered in which the BS and attacker are the players of the game.
In particular, we obtain a \textit{convex-concave} game, and propose an iterative algorithm that converges to a Nash equilibrium (NE) of the game.
This analysis provides an upper bound on the downlink sum-rate of massive MIMO systems under an optimal PC attack.

For the second attack model, the attacker generates jamming signals in both the pilot and downlink data transmission phases.
For this hybrid attack, the attacker is required to have a full-duplex radio.
Specifically, the attacker estimates the channels between users and itself while jamming the uplink pilot transmissions.
These estimates are then used to strengthen the attack during the downlink data transmission phase.
Stochastic optimization techniques are used to find the optimal power allocation at the attacker so as to minimize the downlink sum-rate of the system.

Massive MIMO systems are robust against passive eavesdropping, as the CSI at a legitimate receiver and an eavesdropper are near-orthogonal \cite{Kapetanovic2015}.
However, these systems are vulnerable to an active attacker that contaminates the uplink pilot transmissions.
Therefore, we extend our work in \cite{CNS17_Berk} and analyze the secrecy performance of a massive MIMO system under a PC attack.
Specifically, an attacker receives the information signals intended to users with a much higher signal power by using the PC attack.
We study a problem where the attacker minimizes the maximum of the achievable individual secrecy rates at users.
Our analysis provides an upper bound on the achievable individual secrecy rates in a given massive MIMO system under the PC attack.
Moreover, by introducing chance constraints, we study the case where the attacker does not know the locations of users.
The formulated problems are numerically solved by an iterative method.
Numerical results show that the downlink sum-rate decreases significantly under a PC attack. 
Particularly, when the attacker is close to the BS, the downlink sum-rate of all users is reduced by more than $50\%$.
Another important result of our work is that an attacker without perfect information about user locations is almost as devastating as one with perfect information.
This fact emphasizes the vulnerability of massive MIMO systems to PC attacks. 
Further, we observe that even if the attacker moves farther from the BS, the maximum per-user secrecy rate is reduced by almost $30\%$.

The rest of the paper is organized as follows.
Section \ref{sec:systemodel} describes the system model.
In Section \ref{sec:down_trans_rates}, we compute the downlink transmission rates with/without a PC attack.
Our PC attack under a fixed and optimal BS transmission power is studied in Section \ref{sec:optimal_PC_att}.
We analyze the secrecy rates of the users in massive MIMO systems in Section \ref{sec:secrecy}.
Section \ref{sec:advjam} investigates the hybrid attack model.
We provide numerical results in Section \ref{sec:numerical}, and conclude the paper in Section \ref{sec:conclusion}.

Throughout the paper, we adopt the following notation.
$ \mathbb{E}[\cdot] $ indicates the expectation of a random variable.
Row vectors and matrices are denoted by bold lower-case and upper-case letters, respectively.
$ (\cdot)^{*} $ and  $ (\cdot)^{T} $ represent the complex conjugate transpose and transpose of a vector or matrix, respectively.
Frobenius norm and the absolute value of a real or complex number are denoted by $ \Vert \cdot \Vert $ and $ \vert \cdot \vert $, respectively.
$ \Am \in \mathbb{C}^{M \times N} $ means that $ \Am $ is an $ M \times N $ complex matrix, and $\Id_M$ is an $M \times M$ identity matrix.
$ \mathcal{C} \mathcal{N}(\mu, \sigma^{2}) $ denotes a complex circularly symmetric Gaussian random variable of mean $ \mu $ and variance $ \sigma^{2} $.
$[x]^{+}$ is defined as $\mathrm{max}(x, 0)$.
For simplicity, $\log_2(\cdot)$ is referred to as $\log(\cdot)$.

\section{System Model}
\label{sec:systemodel}

We consider a single-cell massive MIMO system in which the BS (Alice) uses a large array of $M$ antenna elements to transmit/receive independent data streams to/from $K$ single-antenna users (Bobs), where $M \gg K$.
Because of the large $M$, the channel coherence time is not long enough to estimate the CSI of all $M$ downlink channels at each user \cite{Lu2014}.
Therefore, TDD is used instead of FDD (in FDD, the downlink and uplink channels are estimated separately).
In TDD, Alice estimates the CSI for uplink channels after receiving pilot sequences transmitted by Bobs.
If these pilot symbols are not perfectly orthogonal to each other, interference among them causes erroneous channel estimates at the BS.
Assuming channel reciprocity, these estimates are used in setting the precoding matrices of downlink data transmissions.
There is no standardization for massive MIMO systems regarding the orthogonality of the pilot sequences.
However, the authors in \cite{Marzetta2010} suggested assigning an orthogonal time-frequency pilot sequence to each Bob.
Orthogonal space-time block codes can also be utilized, as in 802.11ac systems, to increase the number of orthogonal pilot sequences.
Fig. \ref{fig:pilots} shows an example of eight pilot sequences.
Pilot sequences $\pv_1$ and $\pv_2$ are orthogonal space-time coded sequences.
They are sent in the same time interval ($t_1$) over the same frequency ($f_1$) by two different Bobs.
On the other hand, the orthogonality of $\pv_1$ and $\pv_5$ is guaranteed by transmitting them in different time intervals $t_1$ and $t_2$, e.g., $\pv_1 = 0$ during $t_2$.
Similarly, $\pv_1$ and $\pv_3$ are transmitted on different frequencies $f_1$ and $f_2$, respectively.

The received signal at Alice during the pilot transmission phase is given by:
\begin{figure}[!t]
\centering
\includegraphics[width=65mm]{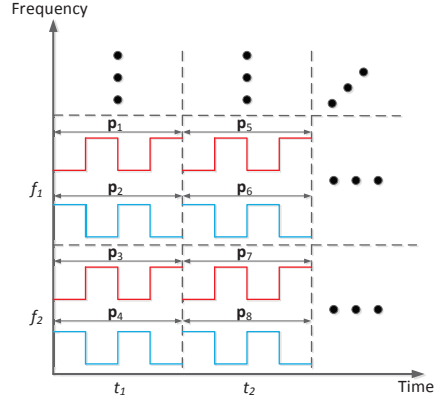}
\caption{Orthogonality of pilot sequences in space-time-frequency domain.}
\label{fig:pilots}
\end{figure}
\begin{align}
\Ym_A = \sum_{i = 1}^{K} \sqrt{P_k}\hv_k^T \pv_k + \Wm
\end{align}
where $\hv_k^T \in \mathbb{C}^{M \times 1}$ represents the uplink channel from Bob$_k$ ($k$th Bob) to Alice.
The $m$th entry, $m \in \{ 1, \cdots, M \}$, of this vector is given by $ h_k^{(m)} = \sqrt{\theta_k}g_k^{(m)}$, where $\theta_k$ and $g_k^{(m)} \sim \mathcal{C} \mathcal{N} (0, 1)$ represent the path-loss component (large-scale fading) and small-scale effects of the channel (Rayleigh fading), respectively.
Note that $\theta_k$  is the same for all antennas, so $\hv_k$ can be written as $ \hv_k = \sqrt{\theta_k} \gv_k$, where $\gv_k$ is a vector of all $g_k^{(m)}$, $m \in \{ 1, \cdots, M \}$.
$\pv_k \in \mathbb{C}^{1 \times L}$ is the transmitted pilot sequence by Bob$_k$, where $L$ is the number of symbols in the pilot sequence.
As these pilot sequences are orthogonal to each other, $\pv_k \pv_l^*  = 0$ $\forall \; k$ and $l \in \mathcal{K}$, where $k \neq l$ and $ \mathcal{K} = \{ 1, \cdots, K \}$.
We assume $\pv_k$ is a unit vector (i.e., $\pv_k \pv_k^* = 1$ $\forall k \in \Kc$).
$P_k$ is the pilot transmission power at Bob$_k$.
$\Wm$ is the additive white Gaussian noise (AWGN) matrix, whose entries are zero-mean, unit-variance normal random variables.

Without loss of generality, consider the estimation of $\hv_i$ at Alice.
Let $\hat{\hv}_i$ represent this estimate.
Under a priori knowledge of $ \pv_i $, Alice post-multiplies the received signal by $ \pv_i^* $ and divides it by $\sqrt{P_i}$ and $L$ to obtain:
\begin{align}
\hat{\hv}_i^T & = \dfrac{\Ym_A \pv_i^*}{\sqrt{P_i} L} = \sum_{k = 1}^{K} \dfrac{\sqrt{P_k}\hv_k^T \pv_k \pv_i^* }{\sqrt{P_iL}} + \dfrac{\Wm \pv_i^* }{\sqrt{P_i}L} \nonumber \\
&= \hv_i^T + \tilde{\wv}_i^T
\label{eq:ch_est_wnoise}
\end{align}
where $\tilde{\wv}_i^T \triangleq \frac{\Wm \pv_i^* }{\sqrt{P_i}L} \sim \mathcal{C} \mathcal{N} (0, \frac{1}{P_i L} \Id_M)$.

\subsection{PC Attack Model}
\label{subsec:attackmodel}
We now describe the first attack model considered in this paper.
The attacker aims to contaminate pilot transmissions by imposing his signal.
We assume that the attacker knows the pilot sequences used by Bobs (generally, pilot sequences are publicly known).
Because the number of orthogonal pilots is limited, after eavesdropping on the channels for some time, the attacker can learn the pilots assigned to various Bobs.
Let $\xv_J \in \mathbb{C}^{1 \times L}$ be the signal generated by the attacker.
After the attack, the received signal at Alice will be modified as follows:
\begin{align}
\Ym_A = \sum_{k = 1}^{K} \sqrt{P_k} \hv_k^T \pv_k + \hv_J^T \xv_J + \Wm
\end{align}
where $\hv_J^T \in \mathbb{C}^{M \times 1}$ represents the channel vector from the attacker to Alice.
In the literature, $\xv_J$ is often designed such that only a single user is targeted by the attacker \cite{Zhou2012, Basciftci}.
This user is selected randomly without any optimization.
More specifically, $\xv_J$ is often set to $\sqrt{P_J} \pv_k$, where $P_J$ is the average jamming power.
In contrast, in our model, we set $\xv_J$ to:
\begin{align}
\xv_J = \sqrt{P_J} \sum_{k = 1}^{K} \sqrt{\alpha_k} \pv_k
\label{eq:attack1}
\end{align}
where $\alpha_k$ is the ratio between the power that the attacker allocates to pilot $\pv_k$ and the average jamming power.
Note that $\sum_{k = 1}^{K} \alpha_k \leq 1$.
The objective of the attacker is to minimize the downlink sum-rate.
Let $R_k$ be the downlink transmission rate at Bob$_k$.
The attacker's goal can be formulated as follows:
\begin{align}
\label{prob:1}
\underset{ 
\begin{subarray} \\
\{ \alpha_k \; \forall k \in \mathcal{K} \} \end{subarray} }{\text{minimize}} 
& \sum_{k \in \mathcal{K}} R_{k}
\end{align}
subject to $\alpha_k \geq 0   \; \forall k \in \mathcal{K}$ and $\sum_{k = 1}^K \alpha_k \leq 1$.

\section{Downlink Transmission Rates}
\label{sec:down_trans_rates}

In this section, we analyze the downlink sum-rate in the underlying massive MIMO system with/without the aforementioned PC attack. 

\subsection{No PC Attack Scenario}

In massive MIMO systems, the BS often applies MRT precoder \cite{Bj2016, Larsson2014, Lu2014, fernandes2013inter}.
For conventional MIMO systems, MRT gives rise to inter-user interference.
However, as $M$ tends to infinity, the channels between the BS and individual users become orthogonal to each other, and they individually reduce to single-input single-output (SISO) channels.
In this case, MRT is the optimal precoder.
Let $s_k$ be the downlink information signal intended to Bob$_k$ $\forall k \in \mathcal{K}$, and let $\vv_k^T \in \mathbb{C}^{M \times 1}$ be its normalized precoder, i.e., $\vv_k \vv_k^* = 1$.
The received signal at Bob$_k$ in the downlink data transmission phase is given by:
\begin{align}
y_k &= \sum_{i = 1}^K \sqrt{P_i^{(d)}} \hv_k \vv_i^T s_i + w_k^{(d)} 
\label{eq:rec_sig_bobk}
\end{align}
where $P_k^{(d)}$ and $w_k^{(d)}$ are, respectively, the allocated power to $s_k$ at Alice and the AWGN with zero-mean and unit-variance at Bob$_k$.
Employing MRT precoding, $\vv_k^T$ is given by $\vv_k^T = (\hat{\hv}_k^* / \Vert \hat{\hv}_k \Vert ) $.
The achievable downlink rate at Bob$_k$ becomes:
\begin{align}
\label{eq:original_rates}
R_k = \log \left( 1 + \frac{ P_k^{(d)} \vert  \hv_k \vv_k^T \vert^2}{ \sum_{l \in \{ \mathcal{K} \setminus k \} }P_l^{(d)} \vert  \hv_k \vv_l^T \vert^2 + 1} \right), \; k \in \mathcal{K}.
\end{align}
Note that the precoding vectors are computed based on channel estimates.

Next, we study the asymptotic behavior of $R_k$ as $M \rightarrow \infty$.
Such asymptotic analysis is needed later on for comparison with the case under a PC attack.
Consider the inter-user interference term $P_l^{(d)} \vert  \hv_k \vv_l^T \vert^2$ in (\ref{eq:original_rates}).
Scaling this term by $M$ and taking the limit as $M \rightarrow \infty$, we end up with:
\begin{align}
\lim_{M \rightarrow \infty} \frac{ P_l^{(d)} \vert \hv_k \vv_l^T \vert^2} {M} = 0
\label{eq:interuser_interference}
\end{align}
$\forall k$ and $l \in \mathcal{K}$, where $k \neq l$ (see Appendix \ref{App:proof_interuser_interference} for the proof).
The underlying intuition behind this result is that entries of small-scale channel components of Bob$_k$ and Bob$_l$ are independent random variables of zero-mean and unit-variance.
Hence, $\lim_{M \rightarrow \infty} \gv_l \gv_k^* / M = 0$.
Similarly, $\lim_{M \rightarrow \infty} \gv_l \tilde{\wv}_k^* / M = 0$.
This is a result of the channel orthogonality in massive MIMO systems.
On the other hand, for the term in the numerator in (\ref{eq:original_rates}), we have
\begin{align}
\lim_{M \rightarrow \infty} \frac{ P_k^{(d)} \vert \hv_k \vv_k^T \vert^2} {M} = \frac{ P_k^{(d)} \theta_k^2}{\theta_k + \frac{1}{P_k L}} > 0
\label{eq:snr_m}
\end{align}
(see Appendix \ref{App:proof_numerator} for the proof).
Hence, the downlink rate at Bob$_k$ behaves asymptotically as:
\begin{align}
R_k \sim \log \left( 1 + \frac{ P_k^{(d)} \theta_k^2}{(\theta_k + \frac{1}{P_k L}) \frac{1}{M}} \right).
\label{eq:sinr_wattack}
\end{align}
In our paper, we consider a finite but sufficiently large $M$, with $M \gg K$, so the channels are near-orthogonal.
As a result, the inter-user interference can be neglected as in (\ref{eq:interuser_interference}).
Moreover, for a sufficiently large $M$, $\vert \hv_k \vv_k^T \vert^2 / M$ approaches the result in (\ref{eq:snr_m}) (\cite{ Lu2014,Marzetta2010, fernandes2013inter}).
(In Section \ref{sec:numerical}, we numerically verify these results.)
As explained before, $\theta_k$ is the large-scale channel components at Bob$_k$.
Equation (\ref{eq:sinr_wattack}) indicates that the SINR does not depend on the small-scale fading components, as they are averaged out by the large antenna array (channel hardening).
The term $(1/M)$ in the equation comes from the AWGN $w_k^{(d)}$ at Bob$_k$.
For example, as $M \rightarrow \infty$, the noise term vanishes and the SINR tends to infinity.
Another noise term arises due to the channel estimation error $\tilde{\wv}_i$.
For example, as the length of the pilots, $L$, increases, the second term in the denominator becomes smaller.
This leads to an increase in the downlink rate.
The same effect is also observed when the power allocated for pilots increases.

In this paper, we consider two different downlink transmit power allocation strategies at Alice: ``fixed'' and ``optimal''.
Both strategies are subject to an average power constraint $P_A$.
Under the fixed power allocation, $P_k^{(d)}$ $\forall k \in \mathcal{K}$ is known to the attacker.
For example, based on a fairness criterion, these values may be determined before the pilot transmission phase (e.g., when Bobs are registered with the network), and Alice may convey this information to Bobs through a feedback channel.
If the attacker eavesdrops on this channel, she can obtain the power allocation values.
In an instance of this setup, Alice may simply allocate powers uniformly to the information signals, i.e., $ P_1^{(d)} = \cdots = P_K^{(d)} = P_A / K $.
On the other hand, under the ``optimal'' power allocation strategy, Alice relies on the well-known water-filling technique to assign powers, using $(\theta_k + (P_k L)^{-1}) / (M \theta_k^2)$ as the water levels (see, e.g., \cite{Tse2005}).

\subsection{Presence of PC Attack}

Under the attack model in (\ref{eq:attack1}), the following channel estimation is performed at Alice for each Bob$_k$:
\begin{align}
\hat{\hv}_k &= \hv_k + \sqrt{ \alpha_k u_k } \hv_J + \tilde{\wv}_k
\end{align}
where $u_k $ is the ratio between the average power at the attacker and the pilot transmission power at Bob$_k$, i.e., $u_k = P_J / P_k $.
In the rest of the paper, we assume that $u_k$ is known to the attacker.
Previously, we assumed that the attacker learns the pilot sequences by eavesdropping on the uplink transmissions.
The attacker can similarly learn the pilot transmission power.
Also note that these transmission powers are fixed in the current cellular systems.
Alice is not aware of the presence of the attacker, so she treats $\hat{\hv}_k$ as the correct channel estimate.
Employing MRT precoding based on this estimation, Alice computes the precoder vector of $s_k$ as:
\begin{align}
\vv_k^T = \frac{( \hv_k + \sqrt{\alpha_k u_k}\hv_J + \tilde{\wv}_k )^*}{ \Vert \hv_k + \sqrt{ \alpha_k u_k }\hv_J + \tilde{\wv}_k \Vert }.
\end{align}
Substituting this precoder vector in (\ref{eq:original_rates}), the attacker's optimization problem in (\ref{prob:1}) becomes non-convex. 
To obtain a tractable attack model, we analyze the asymptotic behavior of $R_k$ as $M \rightarrow \infty $.
Following the same steps as in the case of no attacker, the following expression can be obtained as $M \rightarrow \infty$:
\begin{align}
R_k = \log \left( 1 + \frac{P_k^{(d)} M \theta_k^2}{\theta_k + \alpha_k u_k \theta_J + \frac{1}{P_k L}} \right).
\label{eq:rsumwattack}
\end{align}
As $M$ increases, the massive MIMO system becomes more resilient to PC attacks.
However, the vulnerability of the system against such an attack can be observed in (\ref{eq:rsumwattack}), which shows that the SINR decreases with an increase in the jamming power $\alpha_k u_k$, with the functional form as given therein.

As in the previous section, a fixed or ``optimal'' power allocation strategy can be employed to calculate each $P_k^{(d)}$.
Fixed power allocation is performed exactly as before, whereas ``optimal'' power allocation corresponds to the following strategy.
Let $\phi_k \triangleq \theta_k + \alpha_k u_k \theta_J + \frac{1}{P_k L}$.
Then, Alice tries to maximize $R_{\text{sum}} = \sum_{k=1}^K R_k$ to obtain the ``optimal'' power allocation vector:
\begin{align}
\left[ P_1^{(d)} \cdots P_K^{(d)} \right] &= \underset{ x_k, \forall k \in \mathcal{K} }{ \text{argmax}} \; \sum_{k = 1}^K \log \left( 1 + \frac{x_k M \theta_k^2}{\phi_k} \right)
\end{align}
subject to $\sum_{k = 1}^K P_k^{(d)} \leq P_A$ and $P_k^{(d)} \geq 0$, $\forall k \in \mathcal{K}$.
Because Alice is unaware of the attack, she will not necessarily solve the above problem.
However, our goal is to observe the effect of PC attack, even if Alice employs the least favorable power allocation scheme from the perspective of the attacker.
This way, we can establish an upper-bound on the downlink sum-rate under a PC attack.

\section{Analysis of Optimal PC Attack}
\label{sec:optimal_PC_att}

\subsection{Fixed Power Allocation at Alice}
\label{subsec:uniformPA}

In this section, we study the optimal PC attack strategy.
Our analysis provides a lower bound on the downlink sum-rate under a PC attack for a given power allocation at Alice.
We incorporate (\ref{eq:rsumwattack}) into problem (\ref{prob:1}), considering fixed power allocation for the information signals at Alice:
\begin{align*}
\textbf{P1}:
 \underset{ 
\begin{subarray} \\
\{\alpha_k, \; \forall k \in \mathcal{K} \} \end{subarray} }{\text{minimize}} 
& \sum_{k = 1}^K \log \left( 1 + \frac{P_k^{(d)} M \theta_k^2}{\theta_k + \alpha_k u_k \theta_J + \frac{1}{P_k L}} \right)  \\
 s.t. \quad \quad & \alpha_k \geq 0   \; \forall k \in \mathcal{K}, \quad \sum_{k = 1}^K \alpha_k \leq 1. \nonumber 
\end{align*}
For a given $k \in \mathcal{K}$, we assume that $\theta_k = A z_k^{-\gamma}$, where $A$ is a constant that depends on the operating frequency, transmit and receive antennas, while $\gamma$ and $z_k$ are the path-loss exponent and the distance between Alice and Bob$_k$, respectively.
Similarly, $z_J$ is the distance between Alice and the attacker.
For simplicity, the antennas at Bobs and the attacker are assumed to be identical, so the same $A$ is considered for all of them.
As a result, the objective function of \textbf{P1} is converted to the following one:
\begin{align}
R_{\mathrm{sum}} = \sum_{k=1}^K \log \left( 1 +  \frac{ P_k^{(d)} M A z_k^{-2\gamma}}{ \alpha_k u_k z_J^{-\gamma} + z_k^{-\gamma} + \frac{1}{A P_k L}  } \right)
\
\label{eq:rsum}
\end{align}
Next, we discuss two different scenarios based on the information available to the attacker.

\subsubsection{Perfect Information}

Here, we assume that the attacker has perfect knowledge of the distances between Alice and individual Bobs as well as her own distance to Alice.
Indeed, this is an idealized scenario (from the attacker's point of view), and is merely studied to provide a benchmark for comparison with the case of uncertainty in distances.
\textbf{P1} is a convex programming problem, and we obtain the optimal solution as follows.
\begin{theorem}
\label{theo:solution}
\textbf{P1} has the following closed-form solution:
\begin{align}
\alpha_k = \left[ \dfrac{ \sqrt{A_k ( A_k + 4/ \lambda )} - A_k - 2 B_k}{2} \right]^+ \; \forall k \in \mathcal{K}
\end{align}
where $$ A_k \triangleq \frac{P_k^{(d)} M A z_J^{\gamma}}{ u_k z_k^{2\gamma} } \; \mathrm{and} \; B_k \triangleq \frac{ z_J^{\gamma} }{ u_k z_k^{\gamma} }+ \frac{ z_J^{\gamma} }{ u_k A P_k L }. $$
$\lambda$ is the \textit{Karush-Kuhn-Tucker} (KKT) multiplier and is chosen such that $\sum_{k=1}^K \alpha_k = 1 $.
It can be easily computed by the \textit{bisection} method as $\sum_{k=1}^K \alpha_k $ is a decreasing function of it.
\end{theorem}
\begin{IEEEproof}
See Appendix \ref{App:A}.
\end{IEEEproof}

\subsubsection{Uncertainty in Distances}
\label{sssec:uncerdis}

Suppose that the attacker does not have perfect knowledge about various distances.
Let $Z_k$ and $Z_J$ be random variables (rvs) that correspond to the Alice-Bob$_k$ and Alice-attacker distances, respectively.
In this case, the expected value of $R_{\mathrm{sum}}$ is given by:
\begin{align}
\mathbb{E} [ \; R_{\mathrm{sum}} \;] =&  \mathbb{E} \left[ \; \sum_{k=1}^K \log \left( 1 +  \frac{ P_k^{(d)} M A Z_k^{-2\gamma}}{ \alpha_k u_k Z_J^{-\gamma} + Z_k^{-\gamma} + \frac{1}{A P_k L}  } \right) \; \right] \nonumber \\
=& \sum_{k=1}^K \mathbb{E} \left[ \; \log \left( 1 +  \frac{ P_k^{(d)} M A Z^{-2\gamma}}{ \alpha_k u_k Z_J^{-\gamma} + Z^{-\gamma} + \frac{1}{A P_k L}  } \right) \; \right]
\label{eq:expectation}
\end{align}
where $Z$ is a generic rv that has the same distribution as $Z_k$ for all $k$.
In (\ref{eq:expectation}), the expectation is taken over $Z$ and $Z_J$.
The last equality follows from the assumption that the distributions of the distances between individual Bobs and Alice are identical.
We further assume that Bobs and the attacker are randomly and uniformly located in a circular ring around Alice.
Let $D_{\mathrm{min}}$ and $D_{\mathrm{max}}$ be the minimum and maximum possible distances between Alice and any Bob, respectively.
Hence, the CDF of $Z$ is given by $\Pr[Z \leq x] = (x^2 - D_{\mathrm{min}}^2)/ (D_{\mathrm{max}}^2 - D_{\mathrm{min}}^2)$ where $x \in [D_{\mathrm{min}}, D_{\mathrm{max}}]$.
Accordingly, the PDF of $Z$ is given by $f_Z(x) = 2x / (D_{\mathrm{max}}^2 - D_{\mathrm{min}}^2)$, for $x \in [D_{\mathrm{min}}, D_{\mathrm{max}}]$.

Let $\Phi_k \triangleq \alpha_k u_k Z_J^{-\gamma} + Z^{-\gamma} + \frac{1}{A P_k L}$.
Under fixed downlink power allocation, the optimal PC attack can be formulated by the following stochastic programming problem: 
\begin{align}
\textbf{P2}:
\underset{ 
\begin{subarray} \\
\{\alpha_k \; \forall k \in \mathcal{K}\} \end{subarray} }{\text{minimize}} 
&  \sum_{k = 1}^K \mathbb{E} \left[\; \log \left( 1 +  \frac{ P_k^{(d)} M A Z^{-2\gamma}}{ \Phi_k  } \right) \; \right] \nonumber \\
 s.t. \quad \quad & \alpha_k \geq 0  \; \forall k \in \mathcal{K}, \quad \sum_{k = 1}^K \alpha_k \leq 1. \nonumber
\end{align}
The objective function in \textbf{P2} can be rewritten as:
\begin{align}
\sum_{k = 1}^K  \int_{D_{\mathrm{min}}}^{D_{\mathrm{max}}} \int_{D_{\mathrm{min}}}^{D_{\mathrm{max}}} \dfrac{4xy}{(D_{\mathrm{max}}^2 - D_{\mathrm{min}}^2)^2}  \log (\Psi(x,y)) \; d x \; d y 
\label{eq:integral}
\end{align}
where
$$ \Psi(x,y) \triangleq 1 +  \frac{ P_k^{(d)} M A x^{-2\gamma}}{ \alpha_k u_k y^{-\gamma} + x^{-\gamma} + \frac{1}{A P_k L}  } $$
$\mathrm{for} \; x,y \in [D_{\mathrm{min}},D_{\mathrm{max}} ]$.
This is a convex programming problem, as the objective function and inequality constraints are all convex functions.
The integral in (\ref{eq:integral}) can be approximated by Simpson's Rule for double integrals, and can be solved efficiently by applying the interior point method.
Note that \textbf{P2} need only be solved offline, so the time complexity of this solution method is not a concern.
We also note that although we only study a uniform distribution for the locations of Bobs and the attacker, any arbitrary distribution can be considered.
The integral operation preserves the convexity, so the same steps can be followed to solve \textbf{P2}.
Our numerical results (not shown for brevity) indicate that for typical values of $P_A$, $A$, $K$, and $D_{\mathrm{max}}$, the attacker should target all Bobs by equally allocating its average power to various pilot sequences under uniform power allocation when $u_k = u_l$ $\forall k, l \in \mathcal{K}$.
That is, $\alpha_k = P_J / K $ $\forall k \in \mathcal{K}$.
This is due to the symmetry of Bobs for this special case, as will be discussed in Section \ref{sec:numerical}. 

\subsubsection{Discussion}
\label{sssec:dis}

Let $\zv \triangleq [z_1, \cdots, z_K]$ be the vector of distances from Alice to various Bobs (known to the attacker).
Let $\alphav^*(\zv,z_J) \triangleq [\alpha_1^*(\zv,z_J), \cdots, \alpha_K^*(\zv,z_J) ]$ and $\alphav^* \triangleq [\alpha_1^*, \cdots, \alpha_K^* ]$ be the optimal solutions to \textbf{P1} and \textbf{P2}, respectively. 
In this case, the objective function of \textbf{P2} becomes $\mathbb{E}_{\Zm,Z_J} [ R_{\text{sum}}(\alphav^*)]$, and $\mathbb{E}_{\Zm, Z_J} [ R_{\text{sum}}(\alphav^*(\Zm,Z_J))]$ becomes the expectation of the optimal solution of \textbf{P1} under perfect information, where $\Zm$ is the vector of i.i.d. distances $Z_1, \cdots, Z_K$.
The expectations are taken over the random distances, as previously explained.
The \textit{expected value of perfect information} (EVPI) is defined as follows:
\begin{align}
\text{EVPI} \triangleq \mathbb{E}_{\Zm,Z_J} [ R_{\text{sum}}(\alphav^*)] - \mathbb{E}_{\Zm, Z_J} [ R_{\text{sum}}(\alphav^*(\Zm,Z_J))].
\end{align} 
Note that EVPI is always greater than or equal to zero, as the case with perfect information outperforms the one with uncertainty.
If EVPI is small, the attacker does not gain much by knowing the exact distances.
It can perform attacks almost as powerful as when perfect information is available.
On the other hand, if EVPI is high, the attacker may try to acquire distance information by estimating Bobs' locations relative to its own.
For example, a group of colluding adversaries can employ localization techniques (e.g., RSSI and time-of-arrival) to estimate Alice-to-Bobs distances \cite{He2005, Sayed2005}.
This requires more complex and costly systems at the attacker.
In Section \ref{sec:numerical}, we study the behavior of EVPI.

\subsection{Optimal Power Allocation}
\label{subsec:optimalPA}

In this section, we derive the optimal PC attack strategy when Alice adopts optimal (the least favorable from the perspective of the attacker) power allocation strategy for downlink data transmissions.
Note that Alice is assumed to be unaware of the attack.
Therefore, she cannot customize her power allocation strategy to combat such an attacker.
However, while the attacker tries to minimize the downlink sum-rate, Alice tries to maximize this rate, without knowing about the attack.
This is a \textit{min-max} problem, and its solution is found as follows.  
As seen from (\ref{eq:rsum}), $R_{\text{sum}}$ is a function of $ \Pm^{(d)} \triangleq \left[ P_1^{(d)} \cdots P_K^{(d)} \right]$ and $\alphav \triangleq [\alpha_1, \cdots, \alpha_K]$.
Thus, the problem can be formulated as a \textit{convex-concave} game; for a fixed $\Pm^{(d)}$, $R_{\text{sum}}(\Pm^{(d)}, \alphav)$ is a convex function of $\alphav$, and for a fixed $\alphav$, $R_{\text{sum}}(\Pm^{(d)}, \alphav)$ is a concave function of $\Pm^{(d)}$.
This means that the attacker needs to solve the following game:
\begin{align}
\textbf{P3}:
 \underset{ 
\begin{subarray} \\
\{\alphav \} \end{subarray} }{\text{minimize}} \;
& \left\lbrace \underset{ 
\begin{subarray} \\
\{ \Pm^{(d)}  \} \end{subarray} }{\text{maximize}} \;
 R_{\text{sum}}(\Pm^{(d)}, \alphav) \right\rbrace \nonumber  \\
 s.t. \quad \quad & \alpha_k \geq 0   \; \forall k \in \mathcal{K}, \quad \sum_{k = 1}^K \alpha_k \leq 1 \nonumber \\
&  P_k^{(d)} \geq 0   \; \forall k \in \mathcal{K}, \quad \sum_{k = 1}^K  P_k^{(d)} \leq P_A \nonumber 
\end{align}
Let an optimal solution of this game, or a \textit{saddle point}, be $(\Pm^{(d)*}, \alphav^*)$.
That is (for any possible power allocation $\Pm^{(d)}$),
$$ R_{\text{sum}}(\Pm^{(d)}, \alphav^*) \leq R_{\text{sum}}(\Pm^{(d)*}, \alphav^*) \leq R_{\text{sum}}(\Pm^{(d)*}, \alphav).$$
This relationship shows that an upper-bound on $R_{\text{sum}}(\Pm^{(d)}, \alphav)$ is obtained by solving \textbf{P3}.
For instance, when $\alphav = \alphav^*$, $\Pm^{(d)*}$ maximizes $R_{\text{sum}}(\Pm^{(d)}, \alphav^*)$.
This optimal solution is obtained by a well-known water-filling technique.
Specifically,
\begin{align}
P_k^{(d)*} = \left[ \eta - \frac{ \alpha_k^* u_k z_J^{-\gamma} + z_k^{-\gamma} + \frac{1}{A P_k L} }{ M A z_k^{-2\gamma} } \right]^+
\end{align}
where $\eta$ is a water-filling level chosen such that $\sum_{k=1}^K P_k^{(d)} = P_A$.
$\eta$ can be computed by bisection method as this summation is an increasing function of it.
Similarly, when $\Pm^{(d)} = \Pm^{(d)*}$, $\alphav^*$ minimizes $R_{\text{sum}}(\Pm^{(d)*}, \alphav)$.
The optimal solution of this problem was previously given in Theorem \ref{theo:solution}.
We propose to solve this game by using an iterative \textit{Gauss-Seidel} method.
To do that, we first solve $R_{\text{sum}}(\Pm^{(d)}, \alphav)$ for some initial values of $\alpha_k$, e.g., $\alpha_k = 0 $ $\forall k \in \mathcal{K}$ (initially, there is no PC attack).
Then, the obtained $P_k^{(d)}$ values are used in $R_{\text{sum}}(\Pm^{(d)*}, \alphav)$, and this problem is solved with respect to $\alpha_k$ $\forall k \in \mathcal{K}$ as explained in Theorem \ref{theo:solution}.
After this step, the second iteration starts by solving $R_{\text{sum}}(\Pm^{(d)}, \alphav^*)$ using the new values of $\alpha_k$'s.
As the number of iterations increases, a better approximation for the saddle point is obtained.
We evaluate the number of iterations required to reach the Nash equilibrium of this game, and observe that the algorithm almost always converges after $10$ iterations.
\begin{theorem}
Gauss-Seidel iterations converge when used to solve \textbf{P3}.
\end{theorem}
\begin{IEEEproof}
See Appendix \ref{App:theo3}.
\end{IEEEproof}
Note that the above analysis applies to the case of perfect information where distances are known to the attacker.
It can be easily extended to the case where only the probability distribution of distances is known.
The same steps in Section \ref{sssec:uncerdis} are applied to account for the uncertainty.
In particular, the expectation of $R_{\text{sum}}(\Pm^{(d)}, \alphav)$ over $Z_J$ and $Z_k$'s is considered in the objective function of \textbf{P3}.
The resulting problem is still a convex-concave game that can be solved by the Gauss-Seidel method.
We skip this analysis here due to space limitations.

\section{Secrecy Analysis Under PC Attack}
\label{sec:secrecy}

As we analyzed in the previous sections, channels between Bobs and Alice are near-orthogonal in massive MIMO systems as long as $M \gg K$.
Indeed, as $M \rightarrow \infty$, inter-user interference vanishes in massive MIMO systems.
The same reason also makes massive MIMO systems well-protected against passive eavesdroppers (Eve).
For example, channels of Eve and Bobs are near-orthogonal as well, so the mutual information leakage at Eve is negligible. 
However, we showed the vulnerability of massive MIMO systems against an active attacker that contaminates the pilot transmissions.
So far, we only considered the case where the attacker's objective is to minimize the downlink sum-rate.
PC attack also makes Alice transmit information signals towards the attacker, as the precoding vectors are designed based on the erroneous channel estimates, which are linear combinations of CSI at Bobs as well as at the attacker.
Therefore, the attacker receives the information signals intended to Bobs in the data transmission phase.


As a secrecy metric, we consider the individual secrecy rates of Bobs, which ensure that information leakage to an eavesdropper from each information message vanishes \cite{akgun2017exploiting, chen2017individual}.
Specifically, we study a problem in which the attacker, Eve, aims at minimizing the maximum of the achievable individual secrecy rates at Bobs by leveraging PC attacks.
Given MRT precoding at Alice and the same attack model detailed in Sections \ref{sec:systemodel} and \ref{sec:down_trans_rates}, the received signal at the attacker in the downlink data transmission phase is given by:
\begin{align}
y_{\mathrm{eve}} & = \sum_{i=1}^K \sqrt{P_i^{(d)}} \hv_J \vv_i^T s_i + w_J 
\end{align}
where $w_J$ is the AWGN at Eve, and $\vv_k^T = (\hv_k + \sqrt{\alpha_k u_k} \hv_J + \tilde{\wv}_k)^* / \Vert \hv_k + \sqrt{\alpha_k u_k} \hv_J + \tilde{\wv}_k \Vert$.
The individual information leakage rate of $s_k$ is given by: 
\begin{align}
R_k^e = \log \left( 1 + \dfrac{ P_k^{(d)} \vert \hv_J \vv_k^T \vert^2 }{\sum_{l \in \{\mathcal{K} \setminus k \} } P_l^{(d)} \vert \hv_J \vv_l^T \vert^2 + 1} \right), \; \forall k \in \mathcal{K}
\end{align}
Note that $R_k^e$ is obtained from the mutual information between $s_k$ and $y_{\mathrm{eve}}$ where the all other information signals are interpreted as noise.
Similar to the previous sections, we analyze the asymptotic behavior of $R_k^e$ as $M \rightarrow \infty$.
As a result of this analysis, the following limit is obtained:
\begin{align}
\lim_{M \rightarrow \infty} \frac{ P_k^{(d)} \vert \hv_J \vv_k^T \vert^2} {M} = \dfrac{ P_k^{(d)} \alpha_k u_k \theta_J^2 }{\theta_k + \alpha_k u_k \theta_J + \frac{1}{P_k L}} 
\label{eq:limit_leakage}
\end{align}
(see Appendix \ref{App:proof_leakage} for the derivation of (\ref{eq:limit_leakage})).
Therefore:
\begin{align}
\lim_{M \rightarrow \infty} R_k^e = \log \left( 1 + \dfrac{ \dfrac{ P_k^{(d)} \alpha_k u_k \theta_J^2 }{\theta_k + \alpha_k u_k \theta_J + \dfrac{1}{P_k L}} }{\sum_{l \in \{\mathcal{K} \setminus k \} } \dfrac{ P_l^{(d)} \alpha_l u_l \theta_J^2 }{\theta_l + \alpha_l u_l \theta_J + \frac{1}{P_l L}} }  \right).
\end{align}
Note that $\lim_{M \rightarrow \infty} R_k^e $ is independent of $M$.
As we analyze the asymptotic behavior of the system, $\lim_{M \rightarrow \infty} R_k^e$ is referred to as $R_k^e$ in the rest of the paper.

\subsection{Known Distances at Attacker}

The achievable individual secrecy rate for Bob$_k$ is defined by $[R_k - R_k^e]^+$ \cite{chen2017individual}.
Under the fixed power allocation for the downlink signals at Alice and perfect information at the attacker, the optimal PC attack to minimize the maximum of individual secrecy rates is formulated as follows:
\begin{align*}
\underset{ 
\begin{subarray} \\ 
\{ \alpha_k,  \; \forall k \in \mathcal{K} \}
\end{subarray} }
{\text{minimize}} 
& \quad \max \{ R_1 - R_1^e, \cdots, R_K - R_K^e, 0 \} \\
 s.t. \quad \quad & \alpha_k \geq 0  \; \forall k \in \mathcal{K}, \quad \sum_{k=1}^K \alpha_k \leq 1
\end{align*}
We reformulate this problem by introducing a new decision variable $\nu$, such that  $\nu \geq \max \{ R_1 - R_1^e, \cdots, R_K - R_K^e, 0 \}$.
This is equivalent to $\nu \geq 0$ and $\nu \geq R_k - R_k^e $ $\forall k \in \mathcal{K}$.
The problem is now converted to the following one:
\begin{align*}
\textbf{P4}:
\underset{ 
\begin{subarray} \\ 
\{ \nu, \alpha_k,  \; \forall k \in \mathcal{K} \}
\end{subarray} }
{\text{minimize}} 
& \quad \nu \\
 s.t. \quad \quad & R_k - R_k^e - \nu \leq 0 \; \forall k \in \mathcal{K} \\
& \nu \geq 0, \quad \alpha_k \geq 0  \; \forall k \in \mathcal{K}, \quad \sum_{k=1}^K \alpha_k \leq 1
\end{align*}
Note that the solution of \textbf{P4} provides the tightest upper bound on the achievable individual secrecy rate that can be achieved by any Bob in a given massive MIMO system under an optimal PC attack.
However, due to the interference of the information signals at Eve, the first constraint function in \textbf{P4} is not convex.
This makes the problem intractable.
Let $G_k \triangleq P_k^{(d)} \theta_J$.
Therefore, $R_k$ and $R_k^e$ are given as follows:
\begin{align}
R_k &= \log \left( 1 + \dfrac{A_k}{\alpha_k + B_k} \right), \\
R_k^e &= \log \left( 1 +  \frac{\dfrac{G_k \alpha_k}{\alpha_k + B_k}}{ \sum_{l \neq k}^K \dfrac{G_l \alpha_l}{\alpha_l + B_l} } \right).
\end{align}
Let $U_k$ be an upper bound on $R_k - R_k^e$ such that:
\begin{align}
U_k \triangleq R_k - \log \left( 1 +  \dfrac{G_k \alpha_k}{ (\alpha_k + B_k) I_k } \right)
\end{align}
where $ I_k \triangleq ( \sum_{l \neq k}^K G_l / B_l) $.
Note that the function $ R_k - R_k^e$ is a monotonically increasing function with respect to $\alpha_l$ $\forall l \in \mathcal{K}$, $l \neq k$.
An upper bound of this function is obtained when $\alpha_l = 1$ $\forall l \in \mathcal{K}$, $l \neq k$.
Replacing the first constraint in \textbf{P4} by $U_k - \nu \leq 0$ $\forall k \in \mathcal{K}$ makes the problem tractable, and its solution still provides an upper bound on the achievable individual secrecy rate for any Bob.
Furthermore, the logarithm function can be removed by defining $ \hat{\nu} \triangleq 2^{\nu}$. 
Then, \textbf{P4} becomes:
\begin{align*}
\textbf{P5}:
\underset{ 
\begin{subarray} \\ 
\{ \hat{\nu}, \alpha_k,  \; \forall k \in \mathcal{K} \}
\end{subarray} }
{\text{minimize}} 
& \quad \hat{\nu} \\
 s.t. \quad \quad & \dfrac{I_k (\alpha_k + A_k + B_k)}{\alpha_k(I_k + G_k) + B_k I_k} - \hat{\nu} \leq 0 \; \forall k \in \mathcal{K} \\
& \hat{\nu} \geq 1, \quad \alpha_k \geq 0  \; \forall k \in \mathcal{K}, \quad \sum_{k=1}^K \alpha_k \leq 1
\end{align*}
Let $ f_k(\alpha_k) $ denote $ I_k (\alpha_k + A_k + B_k) /( \alpha_k(I_k + G_k) + B_k I_k ) $.
Then, $f_k(\alpha_k)$ is a monotonically decreasing function with respect to $\alpha_k$.
Thus, we propose the following iterative method to numerically solve \textbf{P5}.
Initially, $\max \{ f_1(\alpha_1), \cdots, f_K(\alpha_K) \}$ is found where $\alpha_k = 0$ $\forall k \in \mathcal{K}$.
WLOG, consider that $f_i(\alpha_i)$ is the maximum. 
Then, $\alpha_i \leftarrow \alpha_i + \delta$ where $\delta$ is a positive real number.
After that, the same process is repeated with the new values of $\alpha_k$'s as long as $\sum_{k=1}^K \alpha_k \leq 1$ and $\hat{\nu} \geq 1$.
In Section \ref{sec:numerical}, we numerically compare two upper bounds that are obtained by \textbf{P4} and \textbf{P5}.

\subsection{Unknown Distances at Attacker}

If the location information of Bobs is not available at the attacker, she cannot guarantee any upper bound on the individual secrecy rates.
Therefore, in this case, we replace the first constraint of \textbf{P4} by a chance constraint, as follows:
\begin{align}
\Pr \{ R_k - R_k^e \geq \nu \} \leq \epsilon \; \forall k \in \mathcal{K}
\label{eq:chance_const_1}
\end{align}  
where $\epsilon \in [0,1]$ is a given parameter.
Note that the randomness in (\ref{eq:chance_const_1}) comes from the distances between Bobs and Alice, $Z_k$ $\forall k \in \mathcal{K}$.
(We assume that the attacker knows her distance to Alice, which is a stationary massive MIMO BS.)  
This constraint guarantees that the probability of achieving an individual secrecy rate that is higher than or equal to $\nu$ is less than or equal to $\epsilon$ at Bobs.
That is, only $\epsilon$ fraction of Bobs can achieve an individual secrecy rate above $\nu$.
As (\ref{eq:chance_const_1}) does not have a closed-form expression, \textbf{P4} is intractable for this case as well.
Therefore, we use a similar bounding method as in the known distances case to make the problem tractable.
Let $\hat{U}_k$ be an upper bound on $R_k - R_k^e$ such that:
\begin{align}
\hat{U}_k \triangleq R_k - \log \left( 1 +  \dfrac{G_k \alpha_k}{ (\alpha_k + B_k) \hat{I}_k } \right)
\end{align}
where
\begin{align}
\hat{I}_k \triangleq \sum_{l \neq k}^K \dfrac{ P_l^{(d)} A u_l z_J^{-2 \gamma} }{ u_l z_J^{-\gamma} + D_{\mathrm{max}}^{- \gamma} + \frac{1}{A P_l L}}.
\end{align}
Note that the function $ R_k - R_k^e$ is a monotonically increasing function with respect to both $\alpha_l$ and $z_l$ $\forall l \in \mathcal{K}$, $l \neq k$.
An upper bound of this function is obtained when $\alpha_l = 1$ and $z_l = D_{ \mathrm{max} }$ $\forall l \in \mathcal{K}$, $l \neq k$.
Thus, the following inequalities are obtained:
\begin{align}
& \Pr \{ R_k - R_k^e \geq \nu \} \leq \Pr \{ \hat{U}_k \geq \nu \}  \\
& = \Pr \{ P_k^{(d)} M A \hat{I}_k Z_k^{-2 \gamma} - (\hat{\nu} - 1) \hat{I}_k Z_k^{-\gamma} \geq  \nonumber \\ 
& \quad \hat{\nu} P_k^{(d)} \alpha_k u_k A z_J^{-2 \gamma} + ( \hat{\nu} - 1 ) I_k (\alpha_k u_k z_J^{ -\gamma } + (A P_k L)^{-1} ) \} \\
& \leq \Pr \{  P_k^{(d)} M A \hat{I}_k Z_k^{-2 \gamma} - (\hat{\nu} - 1) \hat{I}_k Z_k^{-2 \gamma} \geq \nonumber \\ 
& \quad  \hat{\nu} P_k^{(d)} \alpha_k u_k A z_J^{-2 \gamma} + ( \hat{\nu} - 1 ) I_k (\alpha_k u_k z_J^{ -\gamma } + (A P_k L)^{-1} ) \} \\
& = \Pr \left\lbrace Z_k \leq \sqrt[2 \gamma]{ \frac{P_k^{(d)} M A \hat{I}_k - (\hat{\nu} - 1) \hat{I}_k}{ J_k } }  \right\rbrace
\label{eq:prob_upper_bond}
\end{align}
where $ J_k \triangleq \hat{\nu} P_k^{(d)} \alpha_k u_k A z_J^{-2 \gamma} + ( \hat{\nu} - 1 ) I_k (\alpha_k u_k z_J^{ -\gamma } + (A P_k L)^{-1} ) $.
To analyze the chance constraint, we exploit (\ref{eq:prob_upper_bond}), which is the CDF of $Z_k$.
As we stated before, $\Pr[Z_k \leq x] = (x^2 - D_{\mathrm{min}}^2)/ (D_{\mathrm{max}}^2 - D_{\mathrm{min}}^2)$ where $x \in [D_{\mathrm{min}}, D_{\mathrm{max}}]$.
Hence, the chance constraint (\ref{eq:chance_const_1}) is converted to:
\begin{align}
 \frac{P_k^{(d)} M A \hat{I}_k - (\hat{\nu} - 1) \hat{I}_k}{ J_k } \leq (\epsilon (D_{\mathrm{max}}^2 - D_{\mathrm{min}}^2) + D_{\mathrm{min}}^2)^{\gamma}
\end{align}
$\forall k \in \mathcal{K}$.
This is equivalent to:
\begin{align}
\frac{ \hat{I}_k ( P_k^{(d)} M A + 1 + Q (\alpha_k u_k z_J^{ -\gamma } + (A P_k L)^{-1} ) )}{\hat{I}_k + Q ( P_k^{(d)} \alpha_k u_k A z_J^{-2 \gamma} + \hat{I}_k (\alpha_k u_k z_J^{ -\gamma } + (A P_k L)^{-1} ) )} \leq \hat{\nu}
\label{eq:unknown_first_constraint}
\end{align}
$\forall k \in \mathcal{K}$ where $Q = (\epsilon (D_{\mathrm{max}}^2 - D_{\mathrm{min}}^2) + D_{\mathrm{min}}^2)^{\gamma}$.
To find the minimum $\hat{\nu}$ for a given $\epsilon$, the same problem as \textbf{P5} is considered at the attacker after replacing the first constraint by (\ref{eq:unknown_first_constraint}).
Note that the constraint function in (\ref{eq:unknown_first_constraint}) is a monotonically decreasing function with respect to $\alpha_k$.
Therefore, the method that we propose for solving \textbf{P5} in the previous subsection can be used here as well. 

In this paper, we study the problem of minimizing the maximum of individual secrecy rates.
The problem in which the attacker aims at minimizing the sum of the individual secrecy rates could be also solved by following similar steps. 
Particularly, the problem would be similarly reformulated, and the new problem would be a convex optimization problem as well.
Due to space limitations, we omit the results here.

\section{Hybrid Full-duplex Attack}
\label{sec:advjam}
\begin{figure}[!t]
\centering
\subfigure[]{\includegraphics[width = 40mm ]{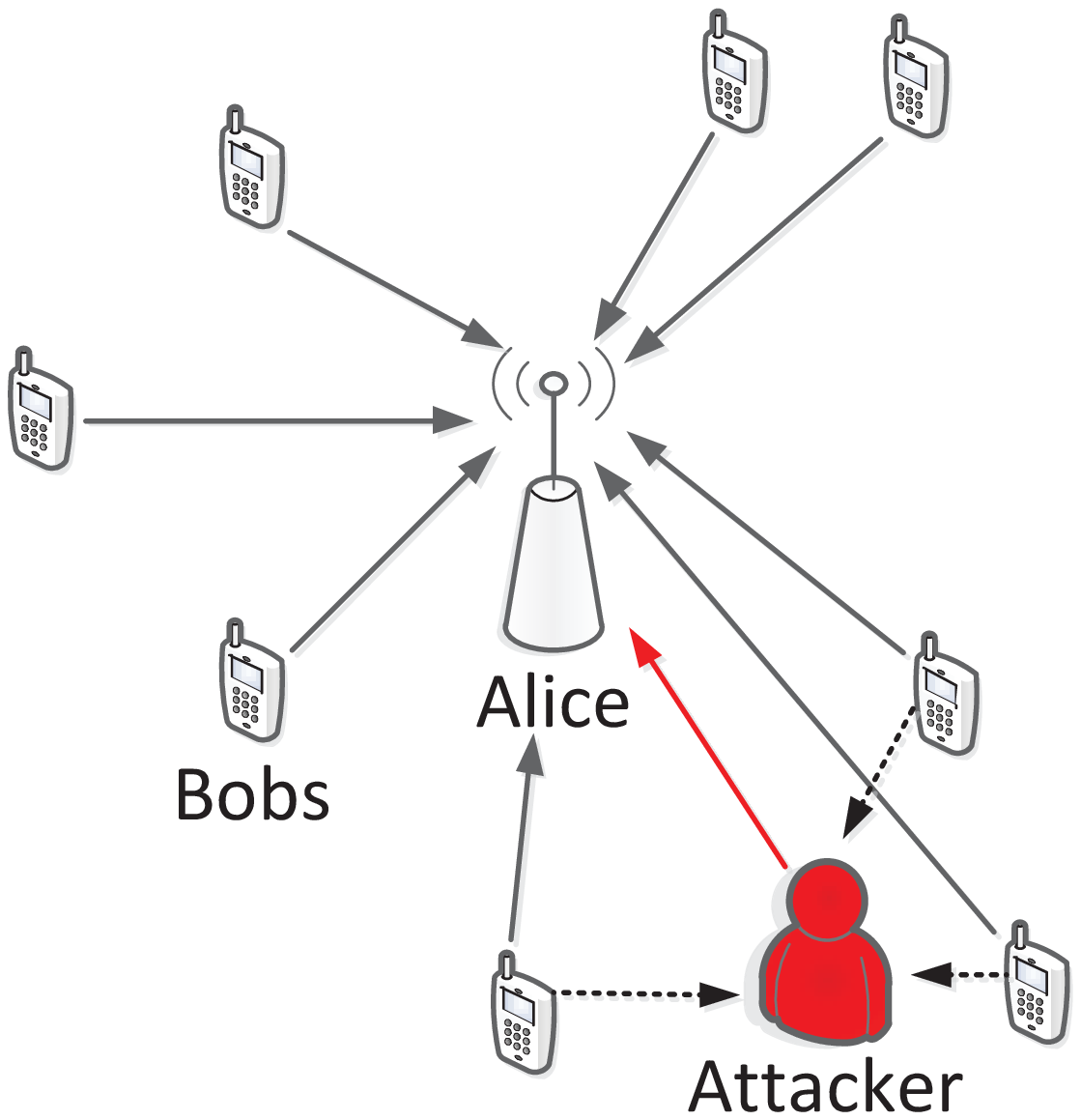}
\label{fig:adv_jam_1}}
\hfil
\subfigure[]{\includegraphics[width= 40mm]{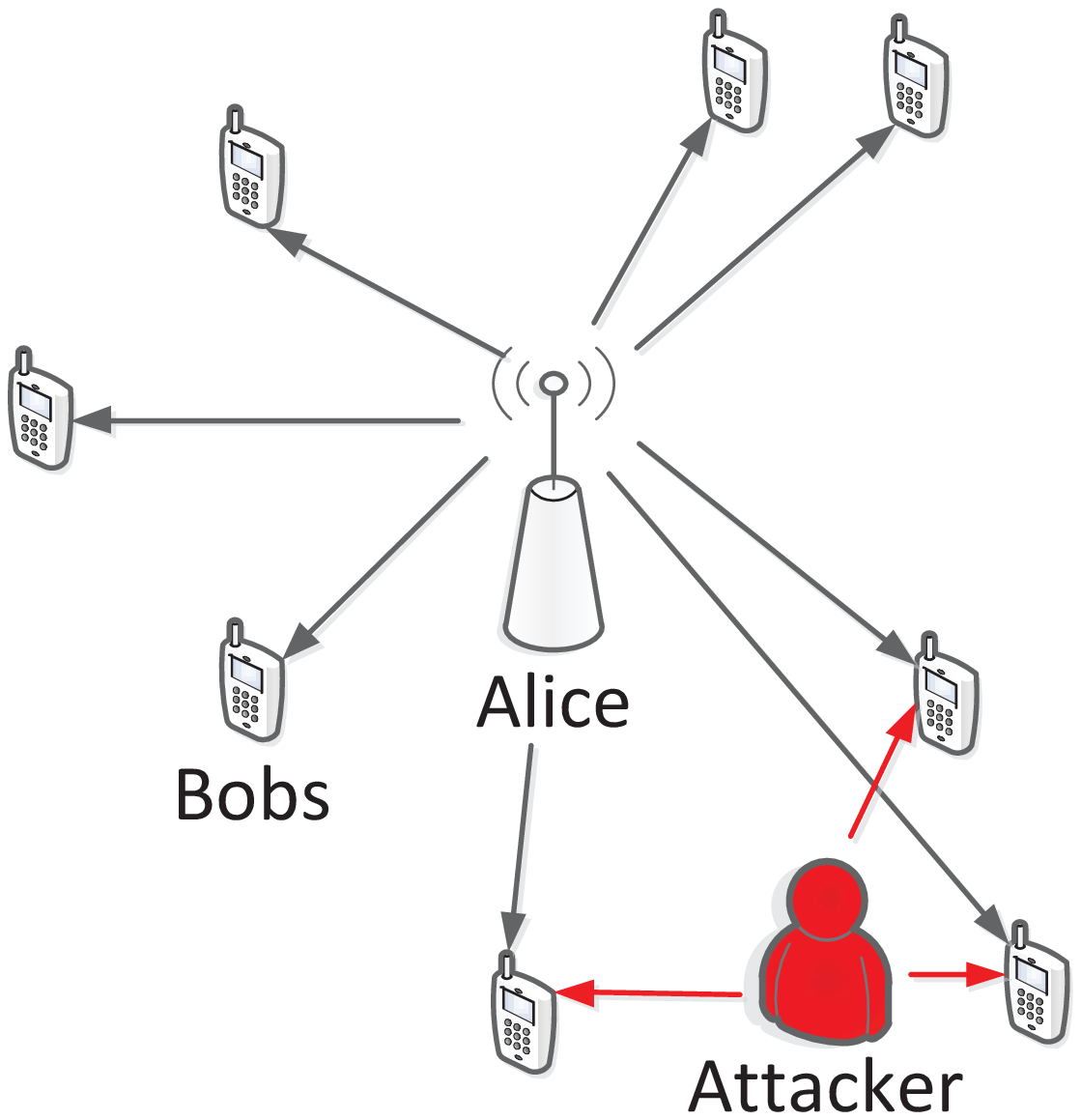}
\label{fig:adv_jam_2}}
\caption{\subref{fig:adv_jam_1} Attacker contaminates the CSI estimation at Alice while overhearing the pilots from Bobs, \subref{fig:adv_jam_2} attacker generates the jamming signals to reduce the signal strength at Bobs during data transmission.}
\label{fig:adv_jam_scenario}
\end{figure}
So far, we have considered jamming the pilot transmission phase.
However, if the attacker is equipped with a full-duplex (FD) radio that allows it to transmit and receive signals simultaneously over the same frequency, a more sophisticated attack can be launched.
Further, a stronger attack can also be launched with a multi-antenna (MIMO) FD-based attacker.
In particular, consider an attacker with an average power constraint over the whole transmission phase (pilot and downlink data phases).
Using an FD radio, the attacker can generate jamming signals during both phases.
For instance, the attacker may contaminate the CSI estimation process at Alice, as in Fig. \ref{fig:adv_jam_1}, without knowing the channels between itself and Bobs.
At the same time, the attacker can overhear the pilots (dashed lines in Fig. \ref{fig:adv_jam_1}) from Bobs using the FD radio, and exploit this knowledge to transmit jamming signals during the downlink transmission phase, as shown in Fig. \ref{fig:adv_jam_2}.
We call this attack a \textit{hybrid} attack, as it combines the PC attack and conventional data-jamming attack. 
Notice that even though the hybrid attack performs at least as good as the PC attack, it requires an additional hardware capability (FD radio) at the attacker.

Even though the attacker needs one antenna to generate a jamming signal in data transmission phase, we study a more general scenario where she is equipped with $N+1$ antennas, where $N > 0 $.
Our goal is to find an optimal strategy for the attacker to minimize the downlink sum-rate, exploiting its multiple antennas.
One of these antennas is reserved for the PC attack, while the others receive the pilot signals from Bobs.
The attacker estimates $\hv_{Jk} \in \mathbb{C}^{1 \times N}$, the channel between Bob$_k$ and herself, during the pilot transmission phase.
The self-interference signal at the receiving antennas of the attacker is canceled by employing FD radio design techniques in \cite{bharadia2013full, bharadia2014full}. 
For example, the self-interference channel is obtained by transmitting a pilot from the antenna that jams the pilot signal.
Then, the self-interference signal is extracted from the received signals using this information.
Let $n_i$ be the $i$th jamming signal in the downlink transmission phase, $ i \in \mathcal{N} = \{1, \cdots, N\}$.
Let $h_{Jk}^{(i)} = g_{Jk}^{(i)} \sqrt{A z_{Jk}^{-\gamma}} $, $i \in \mathcal{N}$ and $k \in \mathcal{K}$, be the channel gain between the $i$th antenna of the attacker and Bob$_k$, where $ z_{Jk}$ denotes the distance between the attacker and Bob$_k$ and $g_{Jk}^{(i)}$ is the small-scale fading. 
$\beta_i$ $\forall i \in \mathcal{N}$ denotes the ratio between the allocated power for $n_i$ and $P_J$.
By using the same PC attack model in Section \ref{subsec:attackmodel} and MRT precoding at Alice, the received signal at Bob$_{k}$ during the downlink data transmission phase is given by:
\begin{align}
y_k = \sum_{i = 1}^K \sqrt{P_i^{(d)}} \hv_k \dfrac{\hat{\hv}_i^*}{ \Vert \hat{\hv}_i \Vert } s_i +
\sum_{i = 1}^N \sqrt{\beta_i P_J} h_{Jk}^{(i)} n_{i} + w_k^{(d)}.
\end{align}
Adding the jamming term to (\ref{eq:sinr_wattack}), the following downlink sum-rate is obtained: 
\begin{align}
\label{eq:rsum_adv}
R_{\mathrm{sum}} = \sum_{k=1}^K \log \left( 1 + \dfrac{ C_k }{ D_k  \left( \sum_{i = 1}^N \beta_i P_J \vert g_{Jk}^{(i)} \vert^2 A z_{Jk}^{-e}  + 1 \right)}  \right)
\end{align}
where $$ C_k \triangleq P_k^{(d)} M A z_k^{-2\gamma} \; \mathrm{and} \; D_k \triangleq \alpha_k u_k z_J^{-\gamma} + z_k^{-\gamma} + \frac{1}{A P_k L} .$$

In this section, we do not analyze the secrecy, and focus on the attack that is studied in Section \ref{sec:optimal_PC_att}.
Given the setup above, we formulate a two-stage stochastic optimization problem to find the optimal attacking strategy that minimizes the downlink sum-rate at Bobs.
This problem can be solved for various scenarios (e.g., perfect information, uncertainity in the distances and channels, etc.) by utilizing the techniques in Section \ref{sec:optimal_PC_att} and the ones presented in this section.
The solutions of these problems are discussed in Section \ref{sec:numerical}.
For now, we explain our solution approach for one of these scenarios.
Specifically, we assume that the distances, powers of information signals, and other constants in \eqref{eq:rsum_adv} are known to the attacker.
In the first stage of the problem, the attacker finds the optimal values of $\alpha_k$ $\forall k \in \mathcal{K}$ without knowing any $g_{Jk}^{(i)}$ $\forall k \in \mathcal{K}$ and $\forall i \in \mathcal{N}$.
In the second stage (after learning $g_{Jk}^{(i)}$ $\forall k \in \mathcal{K}$ and $\forall i \in \mathcal{N}$ during the pilot transmission phase), the attacker optimally allocates the remaining power to the $N$ jamming signals in the data transmission phase, i.e., $\beta_i$ $\forall i \in \mathcal{N}$.
Let $\omega$ represent a certain realization of the channel, $g_{Jk}^{(i)}$, and let $\Omega$ be the set of all realizations.
(Note that $g_{Jk}^{(i)}$ and $\beta_i$ are functions of these realizations.)
Let $t_p$ and $t_d$ be the duration of pilot and data transmission phases, respectively.
The two-stage stochastic problem can be formulated as follows:
\begin{align*}
\textbf{P6}:
\underset{ 
\begin{subarray} \\ 
\{ \alpha_k  \; \forall k \in \mathcal{K} \} \\
\{ \beta_i(\omega) \; \forall i \in \mathcal{N}, \; \forall \omega \in \Omega \} \end{subarray} }
{\text{minimize}} 
& \quad \mathbb{E}_{\omega} \left[ \sum_{k=1}^K \log \left( 1 + \dfrac{ C_k }{ D_k  \left(  E_{k}  + 1 \right)}  \right) \right]  \\
 s.t. \quad & \alpha_k \geq 0  \; \forall k \in \mathcal{K} \\
& \beta_i(\omega) \geq 0  \; \forall i \in \mathcal{N}, \;  \forall \omega \in \Omega \\
&  \frac{F_k}{t_p +  t_d}\leq 1 \;  \forall \omega \in \Omega
\end{align*}
where $F_k \triangleq t_p \sum_{k = 1}^K \alpha_k + t_d \sum_{i = 1}^N \beta_i(\omega)$ and $E_{k} \triangleq  \sum_{i = 1}^N \beta_i(\omega) \vert g_{Jk}^{(i)}(\omega) \vert^2 P_J A z_{Jk}^{-e}$.
Note that $g_{Jk}^{(i)}$ is a continuous random variable.
\textbf{P6} can be approximately solved by creating $T$ realizations, e.g., $\Omega$ has a cardinality of $T$.
In particular, we replace the expectation in \textbf{P6} by the sum of these equiprobable $T$ realizations.
Therefore, we end up with $K$ first-stage decision variables, namely $\alpha_k$ $\forall k \in \mathcal{K}$, and $N T$ second-stage decision variables, namely $\beta_i(\omega)$ $\forall i \in \mathcal{N}$ and $\forall \omega \in \Omega$.
The underlying problem is a convex programming problem, and can be solved by the interior point method.
When $T$ is large (for better approximation), the complexity of solving the problem increases. 
However, as the problem is solved offline, the time complexity is not a concern.

\section{Numerical Results and Discussion}
\label{sec:numerical}

\begin{figure*}[!t]
\centering
\subfigure[]{\includegraphics[width = 52.69mm ]{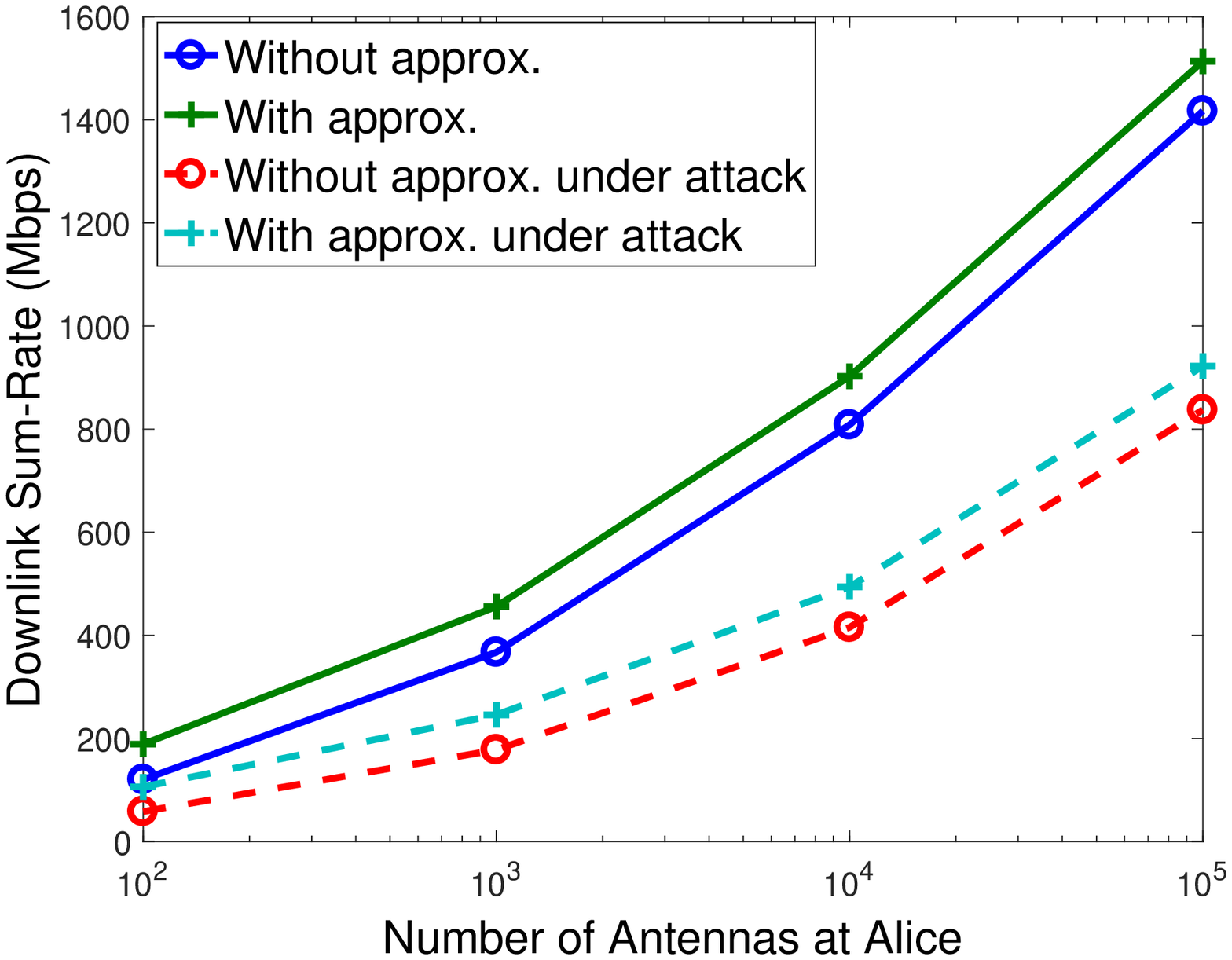}
\label{fig:4a}}
\hfil
\subfigure[]{\includegraphics[width= 52.69mm]{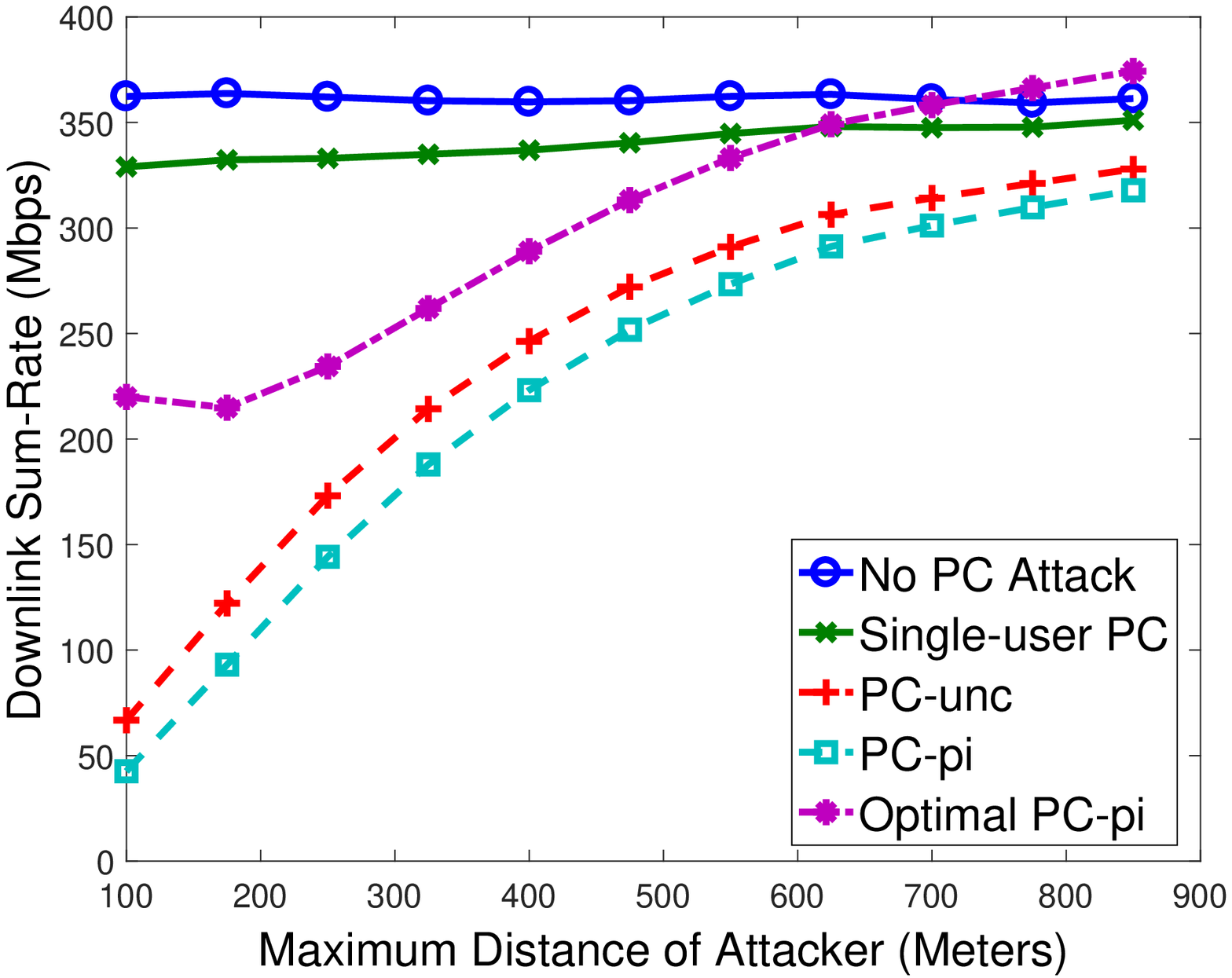}
\label{fig:4b}}
\hfil
\subfigure[]{\includegraphics[width= 52.69mm]{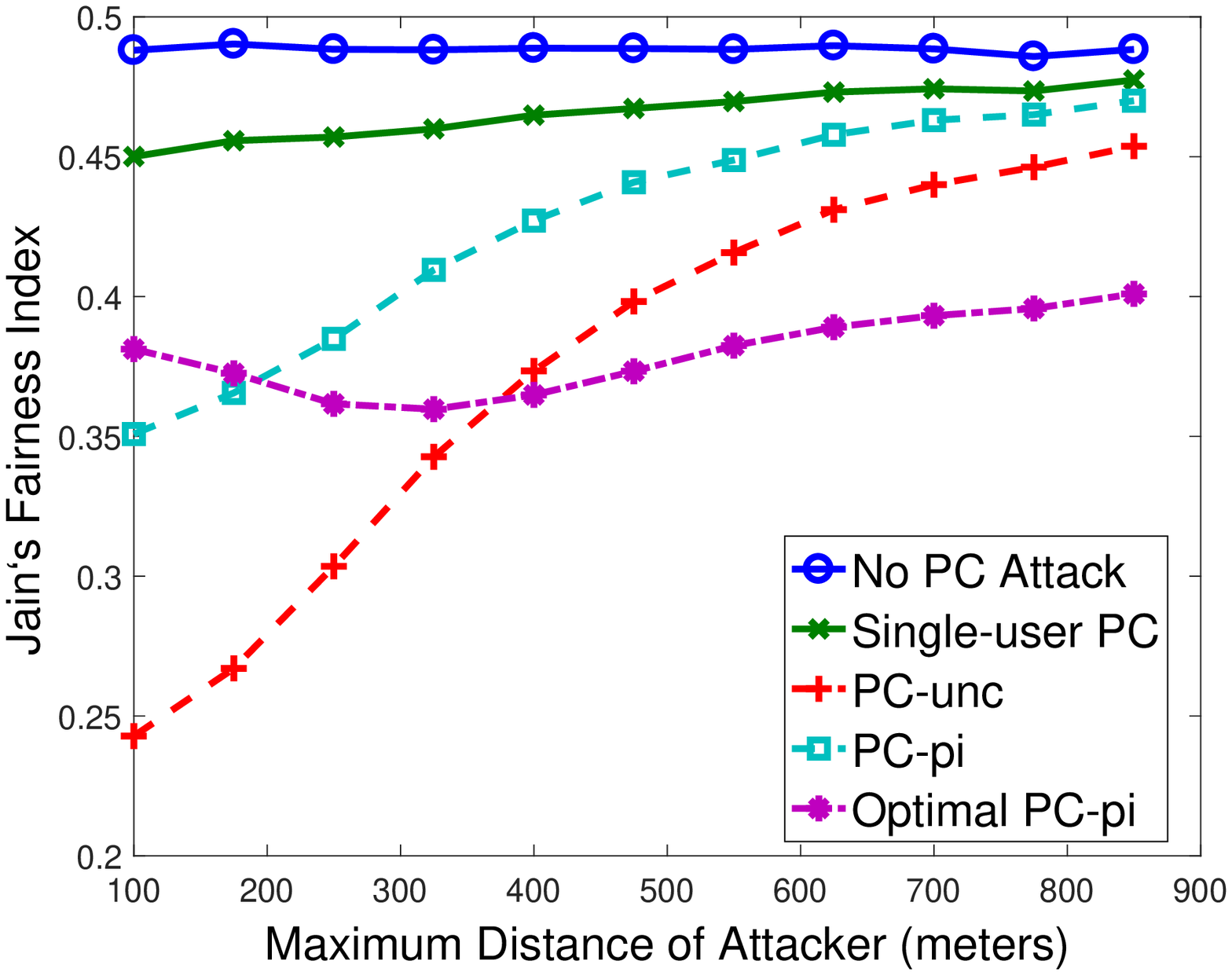}
\label{fig:4c}}
\vfil
\subfigure[]{\includegraphics[width= 52.69mm]{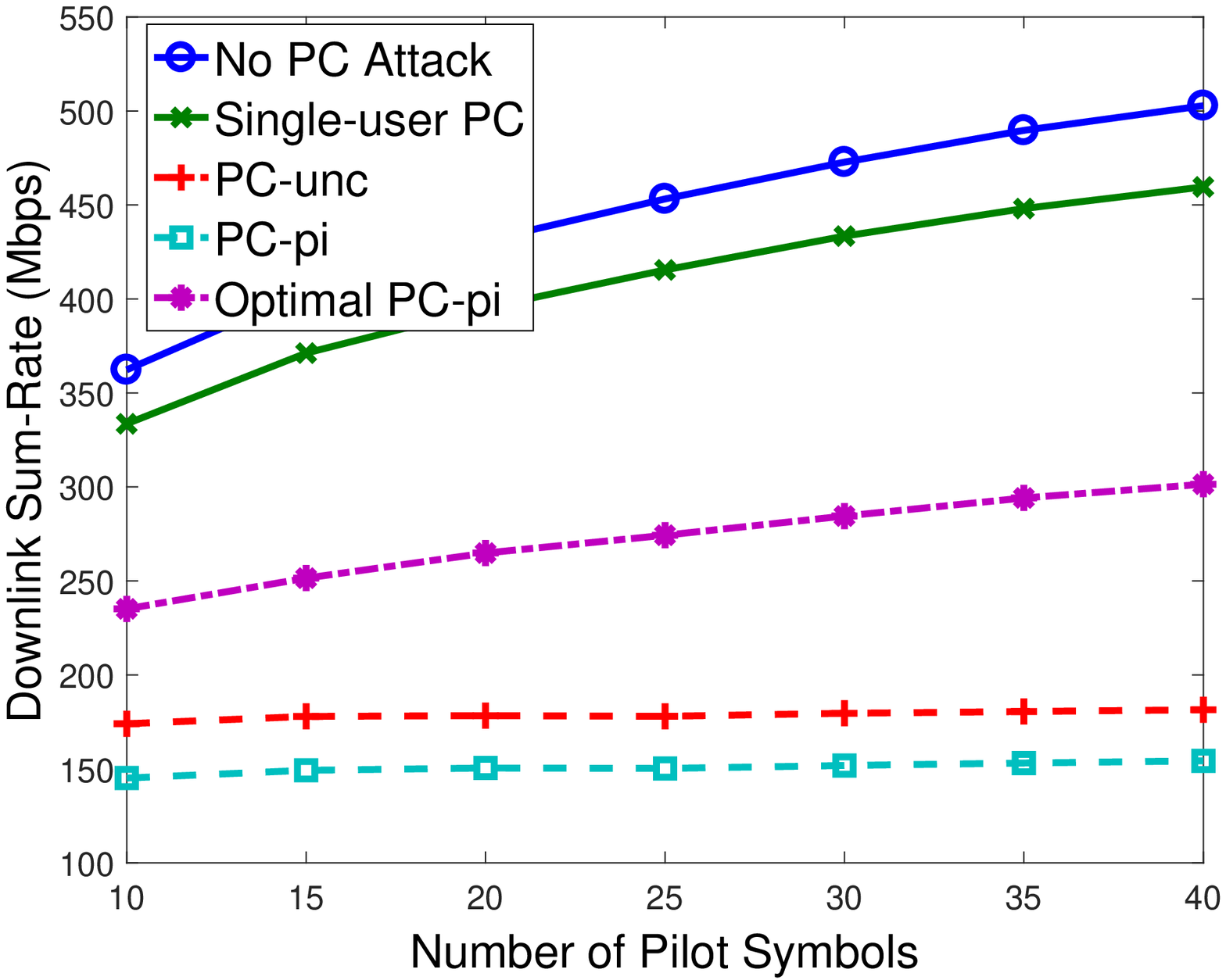}
\label{fig:4d}}
\hfil
\subfigure[]{\includegraphics[width= 52.69mm]{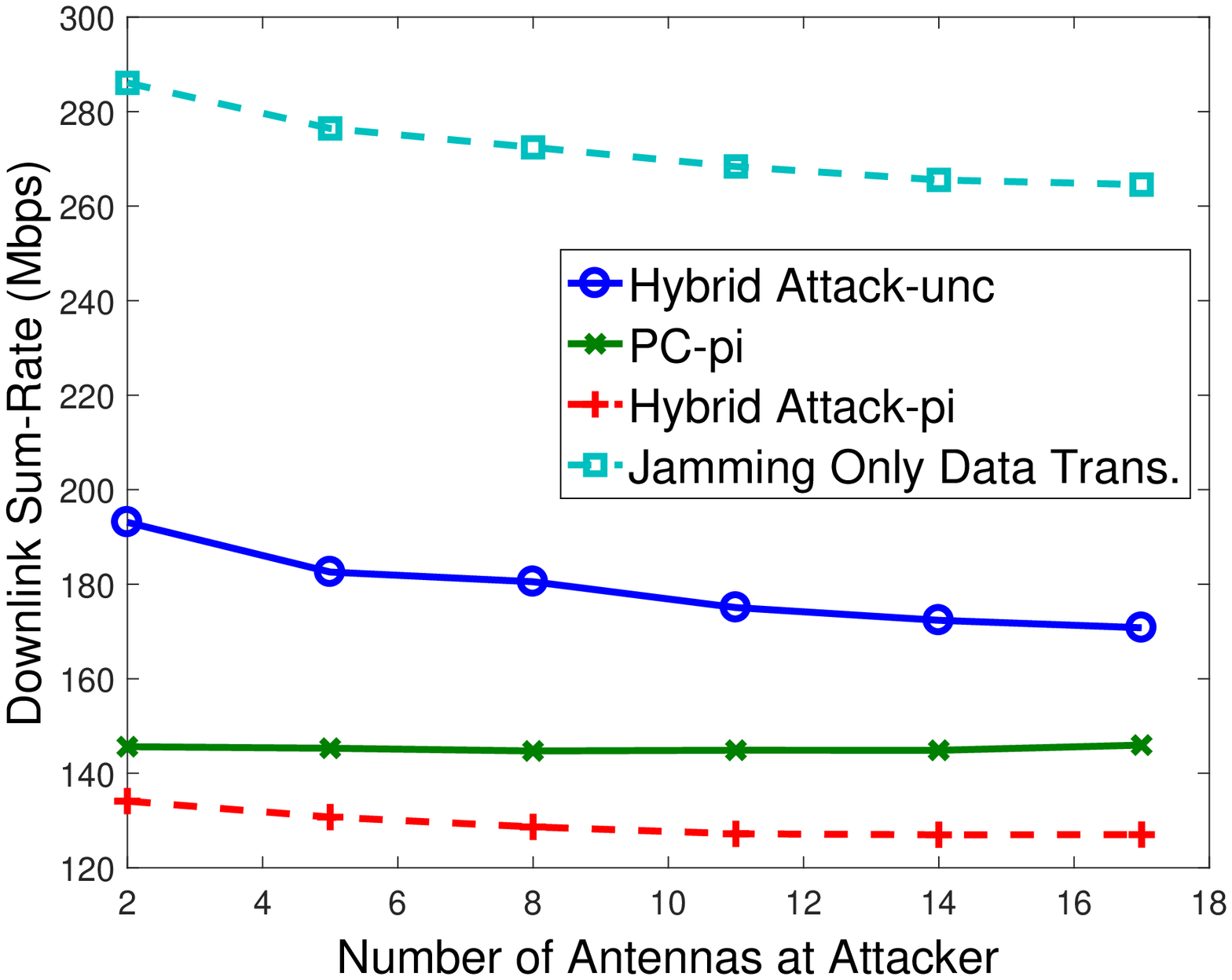}
\label{fig:4e}}
\hfil
\subfigure[]{\includegraphics[width= 52.69mm]{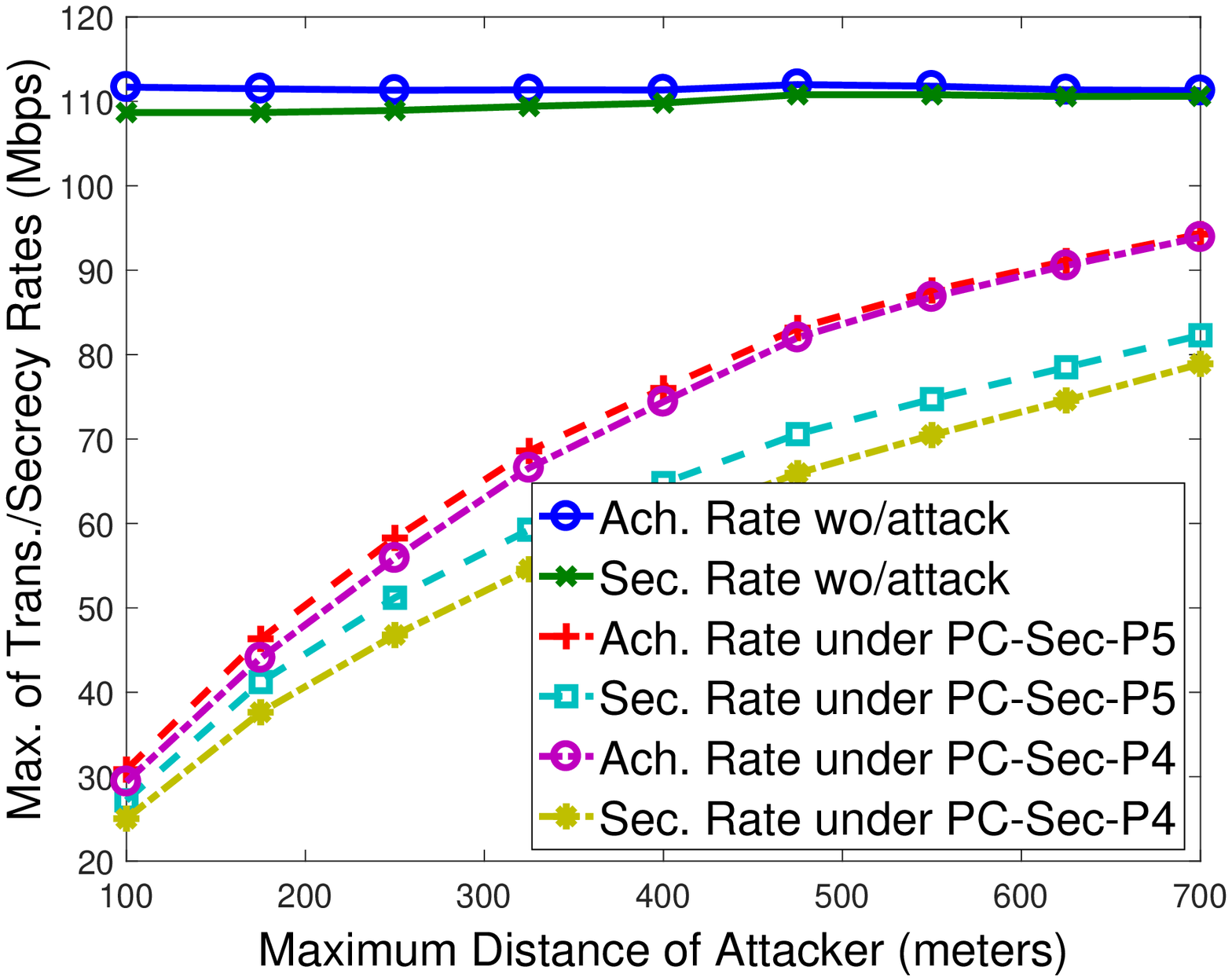}
\label{fig:4f}}
\vfil
\subfigure[]{\includegraphics[width = 52.69mm ]{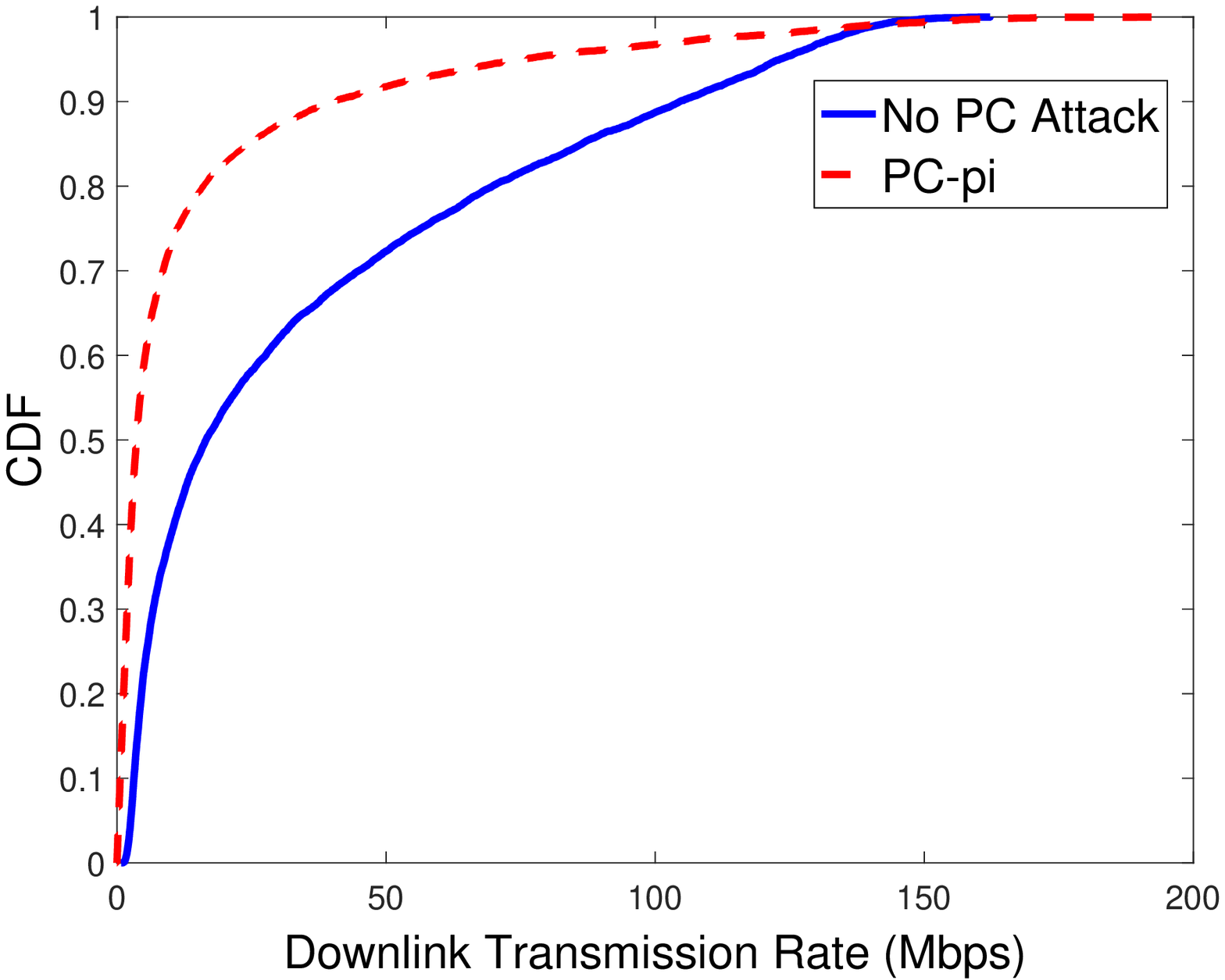}
\label{fig:4g}}
\hfil
\subfigure[]{\includegraphics[width= 52.69mm]{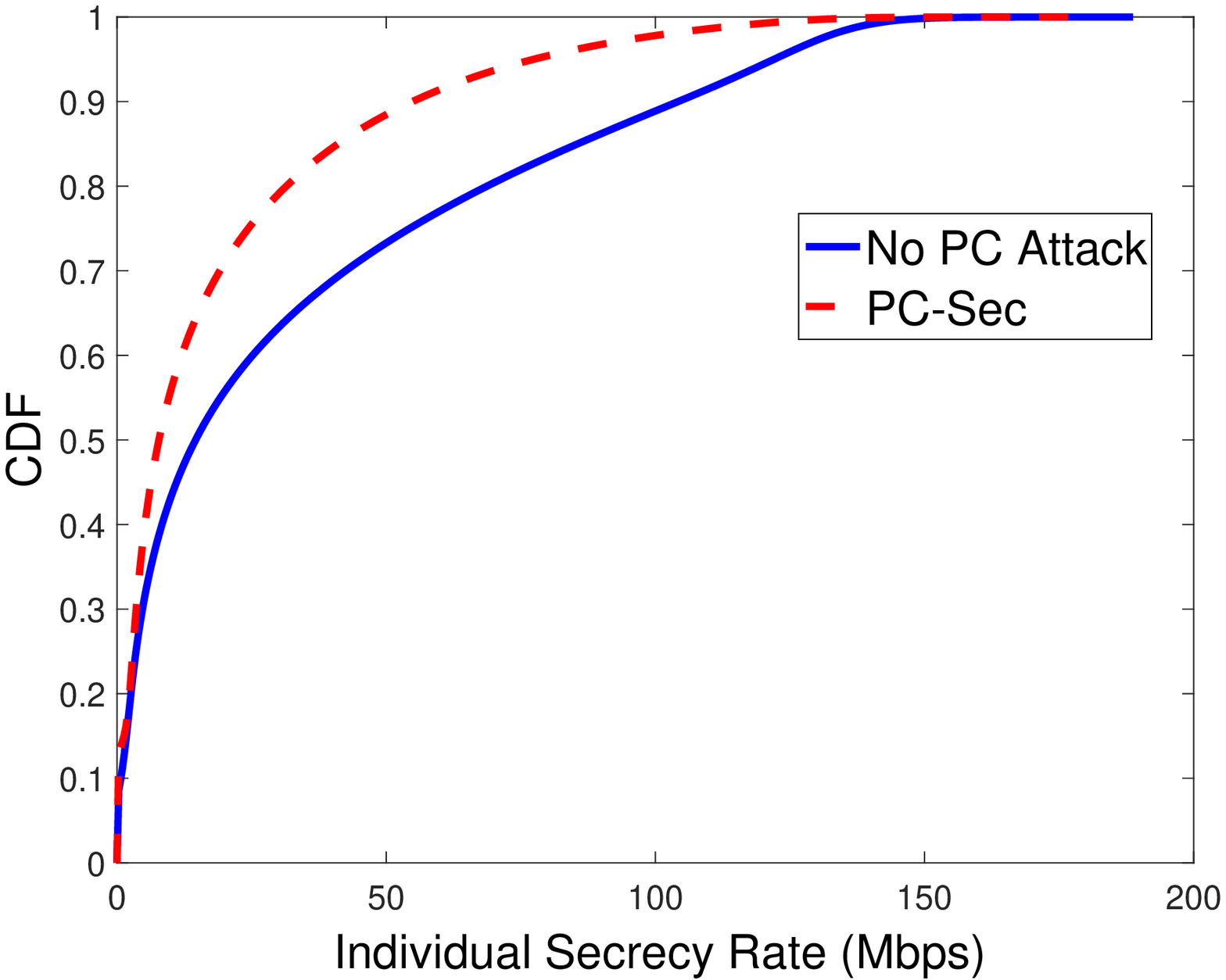}
\label{fig:4h}}
\hfil
\subfigure[]{\includegraphics[width= 52.69mm]{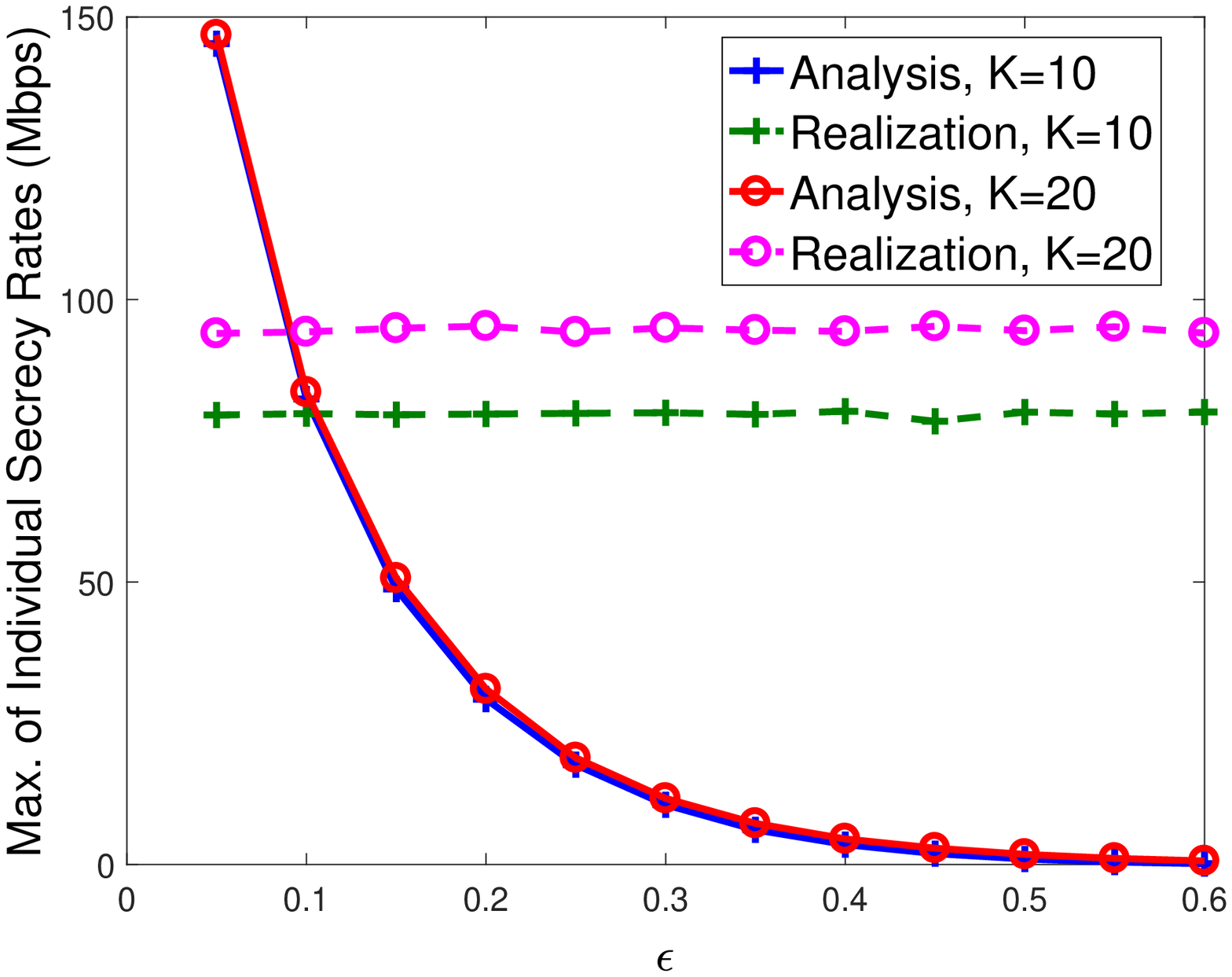}
\label{fig:4i}}
\caption{\subref{fig:4a} Downlink sum-rate vs. $M$ under uniform power allocation at both Alice and the attacker, \subref{fig:4b} downlink sum-rate vs. $D_{\mathrm{max,J}}$, \subref{fig:4c} Jain's fairness index vs. $D_{\mathrm{max,J}}$, \subref{fig:4d} downlink sum-rate vs. number of pilot symbols, \subref{fig:4e}  downlink sum-rate vs. number of antennas at the attacker, \subref{fig:4f} maximum of individual secrecy rates vs. $D_{\mathrm{max,J}}$, \subref{fig:4g} CDF vs. downlink transmission rate, \subref{fig:4h} CDF vs. individual secrecy rate, \subref{fig:4i} maximum of individual secrecy rates vs. $\epsilon$.}
\label{fig:4}
\end{figure*}

We model the channel gain from each transmit antenna to each receive antenna as $h = g \sqrt{A d^{-3.522}}$, where $g \sim \mathcal{C} \mathcal{N} (0,1)$ and $A = 3.0682 \times 10^{-5}$.
The path-loss is modeled using the COST-Hata Model with center frequency is $2$ GHz \cite{abhayawardhana2005comparison}.
The average transmit powers at Alice, Bob$_k$, and the attacker are $46$, $20$, and $30$ dBm, respectively.
The durations of the pilot and data transmission phases are set to be equal \cite{Marzetta2010}.
We consider a $20$ MHz channel with noise floor of $-101$ dBm. 
Bobs and the attacker are uniformly and randomly distributed within a circular ring whose center is Alice and whose outer radius is $D_{\mathrm{max}}$ and $D_{\mathrm{max,J}}$, respectively.
We set $D_{\mathrm{max}}$ to $750$ meters and $D_{\mathrm{min}}$ to $10$ meters.
Our results are averaged over $10^5$ different network realizations. 

We set the number of users $K = 10$.
In Fig. \ref{fig:4a}, we consider uniform power allocation for both the information signals at Alice and the jamming signals at the attacker.
The figure depicts the downlink sum-rate vs. $M$.
It shows that (\ref{eq:sinr_wattack}) and (\ref{eq:rsumwattack}) are good approximations for the downlink rates in (\ref{eq:original_rates}).
Note that the approximation-based sum-rate is slightly higher than the exact values, as the inter-user interference does not perfectly vanish at a finite $M$.
In our subsequent results, we set $M$ to $1000$. 

We observe the effect of the maximum distance between Alice and the attacker ($D_{\mathrm{max,J}}$) in Figs. \ref{fig:4b} and \ref{fig:4c}.
In the case of a single-user PC attack, only one randomly selected Bob is targeted by the attacker.
This attack can also be interpreted as an unintentional interference from a user in an adjacent cell.
It does not have a big impact on the sum-rate.
PC with uncertainty (PC-unc) and PC with perfect information (PC-pi) were explained in Section \ref{subsec:uniformPA}, and optimal PC-pi was studied in Section \ref{subsec:optimalPA}.
Note that optimal PC-pi gives an upper-bound on the sum-rate of a massive MIMO system under an optimal PC attack.
As the attacker moves farther from Alice, the sum-rate increases in all attack schemes.
In Fig. \ref{fig:4b}, EVPI is around $20$ Mbps.
This says that when the attacker knows the distribution of Bobs, it can launch attacks that are almost as powerful as when the attacker has complete CSI.
We also observe that the downlink sum-rate without an attack (no PC attack case) is less than the one with the optimal PC-pi if $D_{\mathrm{max,J}}$ is higher than $700$ meters.
The reason is that Alice uniformly allocates downlink transmission powers in no PC attack scheme, whereas she employs optimal power allocation in the optimal PC-pi.
In Fig. \ref{fig:4c}, we depict Jain's fairness index for different schemes.
Jain's fairness index ranges from $1/K$ to $1$ for the worst and best cases, respectively (if all users have the same downlink rate, the fairness index is $1$).
The figure shows that fairness among Bobs is significantly reduced when PC attacks take place.
PC-unc decreases the fairness more than PC-pi.
The reason behind this phenomena is that when the attacker is close to Alice and knows the distances, Bobs with higher downlink rates are targeted.
Therefore, Bobs are forced to have closer downlink rates, which increases the fairness index.
Note that even though PC-pi makes the fairness higher compared to PC-unc, the sum-rate is lower in PC-pi. 

In Fig \ref{fig:4d}, we set $D_{\mathrm{max,J}}$ to $250$ meters, and study the effect of the number of pilot symbols $L$.
As $L$ increases, the sum-rate increases as well in no PC, single-user PC, and optimal PC-pi attacks.
The reason is that the error in MRT precoding vectors due to erroneous channel estimates decreases, and the signal strength at Bobs increases.
On the other hand, the sum-rate does not increase under the PC-unc and PC-pi attacks.
Note that in these cases, a fixed power is allocated for the information signals at Alice, and she does not exploit the decrease in channel estimation errors.

In Fig. \ref{fig:4e}, we compare hybrid and PC attacks under a similar average jamming power constraint.
We observe that the hybrid attacks outperform PC attacks with respect to the sum-rate.
Moreover, as the number of antennas at the attacker increases, the sum-rate slightly decreases for the hybrid attack.
Note that the hybrid attacks utilizes multiple antennas, whereas PC attacks use a single-antenna.
EVPI for the hybrid attacks is around $60$ Mbps, which is much higher than the one for PC attacks.
The reason is that the hybrid attack includes one more source of uncertainty due to the channels between Bobs and the attacker.
Another important result is that attacking only downlink data transmissions (no jamming during pilot transmission phase) does not have as a great of an impact on performance as the impact of the PC attack.

We evaluate the effect of PC attack on individual secrecy rates in Fig. \ref{fig:4f}.
Specifically, we compare the schemes where there is no PC attack and PC attacks whose objective is to minimize the maximum of the individual secrecy rates (PC-Sec).
PC-Sec-P4 and PC-Sec-P5 denote the results of the problems \textbf{P4} and \textbf{P5}, respectively, with known distances.
Note that even though \textbf{P4} is not a tractable problem, we obtain its results with a brute force method.
For each scheme, we show the results of both individual secrecy rates and transmission rates between Alice and Bobs.
It is observed that the massive MIMO systems are resilient to passive eavesdroppers, as the maximum of transmission/secrecy rates are almost the same without a PC attack.
On the other hand, PC attack decreases the maximum of individual secrecy rates from nearly $110$ Mbps to $55$ Mbps when $D_{\mathrm{max,J}}$ is $325$ meters.
Moreover, we observe that when the attacker moves farther from Alice, PC attack still reduces the maximum of individual secrecy rates by almost $30\%$.
It is also noted that the solution of \textbf{P5} is very close to the solution of \textbf{P4}, which provides the tightest upper bound. 

The empirical CDF of downlink transmission rate and individual secrecy rate under various schemes are shown in Figs. \ref{fig:4g} and \ref{fig:4h}, respectively.
$90\%$ of Bobs achieve a transmission rate less than $40$ Mbps under PC-pi.
In the absence of a PC attack, nearly $33\%$ of Bobs achieve a transmission rate higher than $40$ Mbps.
In Fig. \ref{fig:4h}, we observe that $13\%$ of Bobs have a zero individual secrecy rate under PC-Sec, whereas only $7\%$ fraction of Bobs have a zero individual secrecy rate when there is no PC attack.
Moreover, only $5\%$ of Bobs have a secrecy rate above $75$ Mbps.  

In Fig. \ref{fig:4i}, we evaluate our secrecy analysis with unknown distances at the attacker.
We observe the effect of the designed parameter $\epsilon$ on the maximum of individual secrecy rates for both cases where $K=10$ and $K=20$.
Based on our analysis, $0.1$ fraction of Bobs may achieve an individual secrecy rate higher than $83$ Mbps.
When $K=10$, the maximum of individual secrecy rates is just below this threshold value on average.
On the other hand, when $K = 20$, this threshold value is exceeded almost always as expected.
Note that when $\epsilon = 0.6$, the attacker guarantees that at least $0.4$ fraction of Bobs have zero individual secrecy rate, which emphasizes the vulnerability of a massive MIMO system against a PC attack.

\section{Conclusion}
\label{sec:conclusion}

We considered a single-cell massive MIMO system with several mobile users, and demonstrated vulnerabilities of uplink pilot transmissions against jamming attacks.
Specifically, the attacker generates pilot sequences similar to those of users and contaminates the pilot transmissions to distort channel estimation at the BS.
This PC attack reduces the downlink transmission rates, as the beamforming techniques utilized by the BS heavily depend on accurate CSI estimates.
We formulated an optimization problem from the standpoint of the attacker to minimize the downlink sum-rate.
Both cases when the attacker knows or does not know the distances between the BS and users were considered.
Using (stochastic) optimization and game theory, we derived the optimal attacking strategies when the BS employs either fixed or optimal power allocation for downlink transmissions.
We also analyzed the secrecy rates of the users in massive MIMO systems.
In particular, we showed that even though such systems are robust against a passive eavesdropper, the PC attack significantly reduces the maximum of the individual secrecy rates.
Numerical results showed that the downlink sum-rate is reduced by more than $50\%$ if the average distance between the attacker and the BS is less than the one of the users.
We also observed that even if the attacker does not know the channels and the locations of the users, it can launch powerful attacks as if it has the perfect information.
In this work, we assumed that the BS and users are not aware of the attacker.
An interesting future work is to develop counter algorithms to prevent PC attacks.

\appendices

\section{Proof of Equation (\ref{eq:interuser_interference})}
\label{App:proof_interuser_interference}
\begin{align}
\lim_{M \rightarrow \infty} \frac{ P_l^{(d)} \vert \hv_k \vv_l^T \vert^2} {M} 
= \lim_{M \rightarrow \infty} \frac{ P_l^{(d)} \vert \dfrac{ \hv_k \hat{\hv}_l^* }{M} \vert^2} { \dfrac{\Vert \hat{\hv}_l \Vert^2}{M}}
\label{eq:39}
\end{align}
Let us evaluate the limit of the numerator and denominator separately.
The limit of the denominator is given by: 
\begin{align}
\lim_{M \rightarrow \infty} \dfrac{\Vert \hat{\hv}_l \Vert^2}{M} = \theta_l + \frac{1}{P_l L}
\end{align}
The equality is due to the fact that given a vector $\xv \in \mathbb{C}^{1 \times M}$ with a distribution $\mathcal{CN}(\textbf{0}, c \textbf{I})$, $\lim_{M \rightarrow \infty} \xv \xv^* / M = c$ \cite[Lemma~1]{ fernandes2013inter}. 
We analyze the limit of the numerator as follows:
\begin{align}
\lim_{M \rightarrow \infty} \dfrac{ \hv_k \hat{\hv}_l^* }{M} = \lim_{M \rightarrow \infty} \dfrac{ \sqrt{\theta_k} \gv_k ( \sqrt{\theta_l} \gv_l + \tilde{\wv}_l )^* }{M} = 0
\end{align}
$\gv_k$, $\gv_l$, and $\tilde{\wv}_l$ are independent vectors, and the result follows from \cite[Lemma~1]{ fernandes2013inter}.
The expression in the numerator of (\ref{eq:39}) is a continuous function of $ \hv_k \hat{\hv}_l^* / M$.
Therefore, using the Continuous Mapping Theorem, we have the following result:
\begin{align}
\lim_{M \rightarrow \infty} P_l^{(d)} \vert \dfrac{ \hv_k \hat{\hv}_l^* }{M} \vert^2 = 0
\end{align}
It proves the equation (\ref{eq:interuser_interference}).

\section{Proof of Equation (\ref{eq:snr_m})}
\label{App:proof_numerator}
\begin{align}
\lim_{M \rightarrow \infty} \frac{ P_k^{(d)} \vert \hv_k \vv_k^T \vert^2} {M} &= \lim_{M \rightarrow \infty} \frac{ P_k^{(d)} \vert \dfrac{\hv_k \hat{\hv}_k^* }{M} \vert^2} { \dfrac{\Vert \hat{\hv}_k \Vert^2}{M}}
\end{align}
In this proof, we follow the same steps as in Appendix \ref{App:proof_interuser_interference}.
Therefore, $\lim_{M \rightarrow \infty} \Vert \hat{\hv}_k \Vert^2 / M = \theta_k + 1 / (P_k L)$.
We exploit the Continuous Mapping Theorem to evaluate the limit of the numerator as follows:
\begin{align}
\lim_{M \rightarrow \infty} \dfrac{ \hv_k \hat{\hv}_k^* }{M} &= \lim_{M \rightarrow \infty} \dfrac{ \sqrt{\theta_k} \gv_k ( \sqrt{\theta_k} \gv_k + \tilde{\wv}_k )^* }{M} \\
&= \lim_{M \rightarrow \infty} \dfrac{ \theta_k \gv_k \gv_k^* }{M} = \theta_k
\end{align}
Hence,
\begin{align}
\lim_{M \rightarrow \infty} \frac{ P_k^{(d)} \vert \hv_k \vv_k^T \vert^2} {M} &= \frac{ P_k^{(d)} \theta_k^2} { \theta_k + \frac{1}{P_k L}} 
\end{align}

\section{Proof of Theorem 2}
\label{App:A}

Let us define $$ A_k = \frac{P_k^{(d)} M A z_J^{\gamma}}{ u_k z_k^{2\gamma} } \; \mathrm{and} \; B_k = \frac{ z_J^{\gamma} }{ u_k z_k^{\gamma} }+ \frac{ z_J^{\gamma} }{ u_k A P_k L } $$ $\forall k \in \mathcal{K} $.
Therefore, the objective of \textbf{P1} can be written by
\begin{align}
\label{prob:app1}
R_{\mathrm{sum}} = \sum_{k = 1}^K \log \left( 1 +  \dfrac{ A_k }{ \alpha_k + B_k } \right)
\end{align}
Hence, the Lagrangian function of this problem is given by
\begin{align}
L(\alphav) = \sum_{k=1}^K \log (1 + \dfrac{A_k}{\alpha_k + B_k}) + \lambda ( \sum_{k=1}^K \alpha_k - 1). 
\end{align}
Its first derivative with respect to $\alpha_k$ becomes
\begin{align}
\label{eq:derLag}
\dfrac{\partial L(\alphav)}{\partial \alpha_k} = \dfrac{-A_k}{(\alpha_k + B_k)(\alpha_k + A_k + B_k)} + \lambda.
\end{align}
Let $\alpha_k^*$ $\forall k \in \mathcal{K}$ be the optimal value that minimizes the objective function of \textbf{P1}. 
These values are also the roots of the polynomial functions where the equation (\ref{eq:derLag}) is equal to zero.
Also, note that $\alpha_k^*$ $\forall k \in \mathcal{K}$ is a nonnegative number, and their summation is equal to 1 due to the complementary slackness.
Therefore,
\begin{align}
\alpha_k^* = \left[ \dfrac{ \sqrt{A_k(A_k + 4/\lambda)} -A_k - 2B_k }{2} \right]^+
\end{align}
where $\lambda$ is chosen such that $\sum_{k=1}^K \alpha_k^* = 1$.

\section{Proof of Theorem 3}
\label{App:theo3}

The players of the game described in \textbf{P3} are Alice and the attacker.
In this game, the utility function of Alice is $R_{\mathrm{sum}}(\Pm^{(d)}, \alphav)$, and her strategy is to choose the optimal power allocation for the downlink transmissions.
Similarly, $-R_{\mathrm{sum}}(\Pm^{(d)}, \alphav)$ is the attacker's utility, and her strategy is to find the optimal $\alphav$ to maximize this utility.
The strategy sets of both players are non-empty, compact, and convex subsets of real numbers (the constraints in \textbf{P3} are linear functions).
Furthermore, their utility functions are continuous and diagonally strictly concave.
As a result, the existence and uniqueness of NE is proved for this game, and
Gauss-Seidel method converges to this point \cite{cominetti2012modern}.
 
\section{Proof of Equation (\ref{eq:limit_leakage})}
\label{App:proof_leakage} 
\begin{align}
\lim_{M \rightarrow \infty} \frac{ P_k^{(d)} \vert \hv_J \vv_k^T \vert^2} {M} = \lim_{M \rightarrow \infty} \frac{ P_k^{(d)} \vert \dfrac{\hv_J \hat{\hv}_k^* }{M} \vert^2} { \dfrac{\Vert \hat{\hv}_k \Vert^2}{M}}
\end{align}
Similar to the analysis in Appendices \ref{App:proof_interuser_interference} and \ref{App:proof_numerator}, the limit of the denominator is $\theta_k + \alpha_k u_k \theta_J + \frac{1}{P_k L}$.
We again use the Continuous Mapping Theorem to find the limit of the numerator.
In particular,
\begin{align}
\lim_{M \rightarrow \infty} \dfrac{ \hv_J \hat{\hv}_k^* }{M} &= 
\lim_{M \rightarrow \infty} \dfrac{ \sqrt{\alpha_k u_k} \theta_J \gv_J \gv_J^* }{M} \\
&= \sqrt{\alpha_k u_k} \theta_k
\end{align}
It proves the equation (\ref{eq:limit_leakage}).



\bibliographystyle{IEEEtran}
\bibliography{pilotconRefs}

\end{document}